\documentstyle[emulateapj,pstricks]{article}

\def\Mvir{M_{\rm vir}}
\def\Rvir{R_{\rm vir}}

\def\st{\sigma^2}
\def\Pk{P(k)}
\def\bk{{\bf k}}
\def\dk{\delta_{\bk}}

\def\dc{\delta_{c}}

\def\h1{h^{-1}}
\def\pmk{P_m(k)}
\def\phk{P_h(k)}
\def\LCDM{$\Lambda$CDM}

\def\hMsun{h^{-1}{\ }{\rm M_{\odot}}}
\def\hMpc{h^{-1}{\ }{\rm Mpc}}
\def\ihMpc{h{\ }{\rm Mpc}^{-1}}
\def\hkpc{h^{-1}{\ }{\rm kpc}}
\def\kms{{\ }{\rm km/s}}
\def\Secbanalytic{2}
\def\SecSimulation{3}
\def\SecHalo{4}
\def\SecPSEvol{5.1}
\def\Secddl{5.2}
\def\Secddnl{5.3}
\def\SecDiscussion{6}
\def\SecConclusions{7}

\def\Eqbxi{1}
\def\EqbP{2}
\def\EqbPbxi{4}
\def\EqPS{6}
\def\EqEPS{7}
\def\Eqdh{8}
\def\Eqbnl{9}
\def\Eqbl{10}
\def\Eqbave{11}
\def\Eqmvmax{12}
\def\TabHaloCat{1}
\def\Figpsevol{1}
\def\Figpsevola{1a}
\def\Figpsevolb{1b}
\def\Figpsb5p{2}
\def\Figpsz0{3}
\def\Figb4plin{4}
\def\Figbnonlin{5}
\def\Figbpro{6}
\def\FigclevolI{7}
\def\FigclevolII{8}
\def\Figzmerge{9}

\begin{document}
\slugcomment{{\em Astrophysical Journal, submitted}}
\lefthead{ORIGIN OF BIAS}
\righthead{KRAVTSOV \& KLYPIN}

\title{The origin and evolution of halo bias in linear and non-linear regimes}\vspace{3mm}

\author{Andrey V. Kravtsov and Anatoly A. Klypin}
\affil{Astronomy Department, New Mexico State University, Box 30001, Dept.
4500, Las Cruces, NM 88003-0001}

\begin{abstract}
  We present results from a study of bias and its evolution for galaxy-size
  halos in a large, high-resolution simulation of a low-density, cold dark
  matter model with a cosmological constant. We consider the
  evolution of bias estimated using three different statistics:
  two-point correlation function $b_{\xi}$, power spectrum $b_P$, and a
  direct correlation of smoothed halo and matter overdensity fields
  $b_{\delta}$.  We present accurate estimates of the evolution of
  the matter power spectrum probed deep into the stable clustering regime
  ($k\sim[0.1-200]\ihMpc$ at $z=0$) and find that its shape and
  evolution can be well described, with only a minor modification, by
  the fitting formula of Peacock \& Dodds (1996). The halo power
  spectrum evolves much slower than the power spectrum of matter and
  has a different shape which indicates that the bias is time- and
  scale-dependent. At $z=0$, the halo power spectrum is anti-biased
  ($b_P<1$) with respect to the matter power spectrum at wavenumbers
  $k\sim [0.15-30]\ihMpc$, and provides an excellent match to the power
  spectrum of the APM galaxies at all probed $k$. In particular, both
  the halo and matter power spectra show an inflection at $k\approx
  0.15\ihMpc$, which corresponds to the present-day scale of
  non-linearity and nicely matches the inflection observed in the APM
  power spectrum.  We complement the power spectrum analysis with
  a direct estimate of bias using smoothed halo and matter overdensity
  fields and show that the evolution observed in the simulation in
  linear and mildly non-linear regimes can be well described by the
  analytical model of Mo \& White (1996), {\em if} the distinction
  between formation redshift of halos and observation epoch is
  introduced into the model.  We present arguments and evidence that at
  higher overdensities, the evolution of bias is significantly affected
  by dynamical friction and tidal stripping operating on the satellite
  halos in high-density regions of clusters and groups; we attribute
  the strong anti-bias observed in the halo correlation function and
  power spectrum to these effects.
  
  The results of this study show that despite the apparent complexity,
  the origin and evolution of bias can be understood in terms of the
  processes that drive the formation and evolution of dark matter
  halos. These processes conspire to produce a halo distribution quite
  different from the overall distribution of matter, yet remarkably
  similar to the observed distribution of galaxies.
\end{abstract}


\section{Introduction}


The distribution of galaxies may, in general, be biased with respect to the
overall matter distribution.  Therefore, the galaxy density field can be
used as a probe of matter distribution only if we fully understand how
the galaxy distribution relates to the distribution of matter.
Understanding this relationship, the bias, and its evolution are of primary
importance for the interpretation of the ever increasing amount of
high-quality galaxy clustering data at low and high redshifts. Although
the problem of bias has been studied extensively during the last decade
(e.g., Kaiser 1984; Davis et al. 1985; Bardeen et al. 1986; Dekel \&
Silk 1986; Cole \& Kaiser 1989; Babul \& White 1991), new data on
galaxy clustering at high redshifts (e.g., at $z\lesssim 1$, Le F\`evre
et al.  1996; Shepherd et al. 1997; Carlberg et al.  1997; Connolly et
al.  1998; Postman et al.  1998; Carlberg et al.  1998; and, at $z\sim
3$, Steidel et al. 1998; Giavalisco et al. 1998; Adelberger et al.
1998) and the anticipation of upcoming accurate measurements of galaxy
clustering at $z\approx 0$ from large redshift surveys (e.g., Tegmark
et al. 1998, and references therein) have recently generated
substantial theoretical progress in modelling of galaxy clustering and
bias.

In hierarchical structure formation models, galaxies are assumed to
form inside dark matter (DM) halos via the energy dissipation by
baryons (see Somerville \& Primack 1998 for a recent review).  Mo \&
White (1996) showed how bias of dark matter can be predicted
analytically in the framework of the extended Press-Schechter theory
(Bond et al. 1991; Bower 1991; Lacey \& Cole 1993; see \S
{\Secbanalytic}). This analytical model is rapidly gaining popularity
in theoretical analyses (see, e.g., recent studies by Kauffman, Nusser
\& Steinmetz 1997, Moscardini et al.  1998, and Baugh et al. 1998)
which requires it to be tested and its capabilities and limitations
evaluated.  The model has been tested against numerical simulations by
Mo, Jing \& White (1996), Catelan, Matarrese \& Porciani (1998), Jing
(1998), and Porciani, Catelan \& Lacey (1998), and we present
additional tests in this paper. More elaborate analytical models have
been developed by Catelan et al. (1998ab), Porciani et al. (1998a), and
Sheth \& Lemson (1998).

The effects of non-linearity of the bias on the observable statistics
have been studied by Fry \& Gazta\~naga (1993), Coles (1993), and Mann,
Peacock \& Heavens (1997) using heuristic models of local non-linear
bias.  Recently, Coles (1993) and Dekel \& Lahav (1998) have developed
a formalism for studies of galaxy biasing that allows one to account
explicitly for the non-linearity and stochasticity of the bias. They
have also analyzed the effects of non-linearity and stochasticity on
results of some of the observational analyses. Sherrer \& Weinberg
(1998) have analyzed effects of stochasticity of the local bias on the
correlation function and power spectrum and concluded that
stochasticity should not affect the shape of these statistics in the linear
regime (or, in other words, that in linear regime the bias should be
linear). Most recently, Narayanan, Berlind \& Weinberg (1998) studied 
effects of non-linearity, stochasticity, and non-locality of the
bias using heuristic models applied to large $N$-body simulations. 
From the observational perspective, Pen (1998) showed how the
stochasticity of the galaxy bias can be tested and measured using
redshift-space distorsions, and Tegmark \& Bromley (1998) presented
evidence that present-day galaxy bias is non-linear and stochastic
based on analysis of galaxy clustering in the Las Campanas Redshift
Survey.

The evolution of bias in the linear regime has been recently analyzed by
Tegmark \& Peebles (1998), who generalized the results of Fry (1996)
for the case of stochastic bias in an arbitrary cosmology. Taruya,
Koyama \& Soda (1998) and Taruya \& Soda (1998) used perturbative
analysis to extend the linear analysis of bias evolution into the weakly
non-linear regime. Evolution of halo clustering and bias in the non-linear 
regime, has been analyzed in several recent numerical studies employing
dissipationless simulations (e.g., Brainerd \& Villumsen 1994;
Bagla 1998; Col\'{\i}n et al. 1998; Ma 1999), simulations that include
both dissipationless dark matter and dissipative baryonic components
(e.g., Katz, Hernquist \& Weinberg 1998; Blanton et al.  1998; Cen \&
Ostriker 1998; Jenkins et al. 1998b), and ``hybrid'' studies in which
dissipationless simulations are complemented with a semi-analytical
model of galaxy formation (Roukema et al. 1997; Kauffman et al. 1998ab;
Diaferio et al. 1998; Benson et al. 1998; Baugh et al. 1998; Kolatt et
al. 1998). All of these studies, though very diverse in their methods,
qualitatively agree on one important result: the galaxy bias is
expected to be non-linear, to depend on the properties of the DM halos
and the galaxies they host, and to be a (generally non-monotonic)
function of cosmic time.

In this paper we present results of a detailed analysis of halo
clustering evolution in a large high-resolution dissipationless
simulation of a representative and fairly successful variant of the
cold dark matter (CDM) models: the low-density spatially flat CDM model
with the cosmological constant ({\LCDM}). The primary goal of this
analysis is to identify and study the main processes that drive the
evolution of halo bias in linear and non-linear regimes. Understanding
what causes the bias and particular features of its evolution (notably,
the anti-bias at late stages of evolution in some of the models) is
crucial for the interpretation of current observational and theoretical
results. We focus therefore on the interpretation of general features
of bias evolution observed in our simulations and in studies done by
other authors. The main novel feature of this study is inclusion of
satellite halos located inside virial radii of more massive isolated
halos in the halo catalogs; we refer reader to Col\'{\i}n et al. (1998)
for a more detailed description of our approach.

The approach that we have adopted in this project is to consider bias
estimated using three different statistics: two-point correlation
function $\xi(r)$, power spectrum $P(k)$ (Peebles 1980), and direct
comparison of the smoothed halo and matter overdensity ($\delta$)
fields:
\begin{equation}
b_{\xi}(r)\equiv\sqrt{\xi_h(r)/\xi_m(r)},
\end{equation}
\begin{equation}
b_P(k)\equiv\sqrt{P_h(k)/P_m(k)},
\end{equation}
\begin{equation}
b_{\delta}\equiv \delta_h^R/\delta_m^R,
\end{equation}
where quantities with subscripts $h$ and $m$ correspond to statistics
of the halo and matter distributions, respectively, and superscript $R$
indicates the density fields smoothed on a scale $R$. The three estimates
of bias given above are, of course, related. In the special case of
deterministic local linear bias: $b_{\xi}=b_P=b_{\delta}$.
Nevertheless, in the general case of stochastic non-linear bias, these
estimates are complementary to each other and we have chosen to
consider all three of them to illustrate the manifestations and
properties of the bias. All three functions, $b_{\xi}$, $b_P$, and
$b_{\delta}$, may depend on a number of (both local and non-local)
parameters; the most important point to notice, however, is that they
are different functions of their parameters, and we use the subscripts
to indicate this explicitly. For example, $b_{\xi}$ and $b_P$ have
different scale-dependence because $\xi(r)$ and $P(k)$ are
related via the Fourier transform, $\xi(r)\propto \int P(k)e^{i{\bf
    k}\cdot{\bf r}}d^3k$, which gives:
\begin{equation}
b_{\xi}(r)=\frac{\int b_P(k)P_m(k)e^{i{\bf k}\cdot{\bf r}}d^3k}
{\int P_m(k)e^{i{\bf k}\cdot{\bf r}}d^3k}. 
\end{equation}
The bias $b_{\xi}(r)$ that is scale-dependent in a narrow range of $r$
will be scale-dependent in a wide range of $k$ (see Coles 1993 for
a more detailed discussion). We will show below that the anti-bias
required at small $r$ for the {\LCDM} model to be consistent with the
$z=0$ galaxy correlation function, is also perfectly consistent with
$b_P(k)< 1$ at small wavenumbers (down to $k\sim 0.2\ihMpc$) required
for the model to be consistent with the galaxy power spectrum (Gazta\~naga
\& Baugh 1998; Hoyle et al. 1998).

The paper is organized as follows. In \S {\Secbanalytic}, we give a
brief account of analytical model of the bias developed by Mo \& White
(1996) and how the epochs of halo formation and observation can be separated
in this model (e.g., Moscardini et al. 1998; Catelan et al. 1998a). In \S
{\SecSimulation} and \S {\SecHalo} we describe the numerical simulation
and halo identification and selection algorithms used in our
analysis. We complement the analysis of the evolution of the halo two-point
correlation function based on this simulation and presented in
Col\'{\i}n et al. (1998) with the analysis of ethe volution of 
the halo power spectrum and bias $b_P$ in \S {\SecPSEvol}. 
In \S {\Secddl} and \S {\Secddnl} we present the analysis of the 
evolution of $b_{\delta}$, estimated using smoothed halo and matter 
density fields at different epochs, and identify the
processes that drive this evolution. We discuss the results in \S
{\SecDiscussion} and summarize our conclusions and arguments in \S
{\SecConclusions}.

\section{Analytical model of bias}

An overdensity field $\delta({\bf x})\equiv[\rho({\bf
  x})-\bar{\rho}]/{\bar{\rho}}$ can be quantified in terms of its power
spectrum $\Pk=\langle\vert\dk\vert^2\rangle$, where $\dk$ is the
Fourier transform of $\delta({\bf x})$. Similarly, if the overdensity
field is smoothed on scale $R$ with a spherically symmetric filter
$W(r,R)$, the smoothed field can be characterized by its variance
\begin{equation}
\sigma^2(R)=\langle[\delta({\bf x},R)]^2\rangle=\frac{1}{(2\pi)^3}\int
P(k)\hat{W}(R)^2d^3k, 
\end{equation}
where $\hat{W}$ is the Fourier transform of the window function. In the
following and throughout the paper we use the {\em real space top-hat}
filter: $W(r,R)=(4\pi R^3/3)^{-1}\Theta(R-r)$, where $\Theta(R-r)$ is
the step function. For this filter the scale $R$ can be interchanged
with the mass contained within radius $R$: $M=(4\pi/3)\bar{\rho}R^3$,
where $\bar{\rho}$ is the mean density of the universe. The standard
Press-Schechter model (PS; Press \& Schechter 1974) assumes that any
region of initial comoving size $R$ (or mass $M$) becomes a part of a
virialized halo by redshift $z_f$, if its overdensity extrapolated
linearly to the present epoch is greater than $\delta_c/D_+(z_f)$,
where $D_+(z)$ is the linear growth factor (e.g., Peebles 1980)
normalized to unity at the present epoch.  The value of $\delta_c$ is
motivated by the top-hat collapse model; we use $\delta_c=1.69$
throughout this work.  If the initial overdensity field is gaussian,
the PS model leads to the following expression for the number density
of collapsed halos of mass $M$ at redshift $z$:
\begin{eqnarray}
n(M,z,z_f)dM&=&\frac{1}{\sqrt{2\pi}}\frac{\bar{\rho}}{M}\frac{\dc(z,z_f)}{\sigma^3(M,z)}
\left\vert\frac{d\st(M,z)}{dM}\right\vert\times\nonumber\\
&&\exp\left[-\frac{\dc(z,z_f)^2}{2\sigma^2(M,z)}\right]dM,
\end{eqnarray}
where $\dc(z,z_f)\equiv \dc D_+(z)/D_+(z_f)$ and $\sigma(M,z)$ is the
variance of the initial density field smoothed on scale $M$ and
extrapolated linearly to the epoch $z$. For the reasons that will be
discussed below, we follow Catelan et al. (1998a) in distinguishing
the epoch $z_f$ from the ``observation'' epoch $z$ ($z<z_f$). The $z_f$- and
$z$-dependencies are shown explicitly in the above expression for the
mass function. Note, however, that $n(M,z,z_f)$ does not change with
$z$: for a given power spectrum, the mass function depends only on halo
mass $M$ and $z_f$. Equation ({\EqPS}) translates into the commonly used 
form if we assume $z_f=z$. 

In the seminal paper, Mo \& White (1996; hereafter MW) showed how the
extended Press-Schechter formalism (EPS; Bond et al. 1991; Bower 1991;
Lacey \& Cole 1993) can be used to derive the analytical expression for
the bias of DM halos. The EPS can be used to derive the expression for
the {\em conditional mass function} of halos (Bond et al. 1991).
Namely, the number density of halos of mass $M$ that collapse at epoch
$z_f<z$ in a region of initial Lagrangian radius $R_0$ (mass $M_0$) {\em
  and} the initial overdensity in the growing mode extrapolated linearly to
the present $\delta_0$ ($\delta_0(z)=\delta_0 D_+(z)$) is given by
\begin{eqnarray}
n(M,z,z_f\vert M_0,\delta_0)dM&=&\frac{1}{\sqrt{2\pi}}
\frac{\bar{\rho}}{M}
\frac{\dc-\delta_0}{[\sigma^2-\sigma^2_0]^{3/2}}
\left\vert\frac{d\st}{dM}\right\vert\times\nonumber\\ 
&&\exp\left\{-\frac{[\dc-\delta_0]^2}{2[\sigma^2
    -\sigma^2_0]}\right\}dM,
\end{eqnarray}
where $\delta_c\equiv\delta_c(z,z_f)$, $\delta_0=\delta_0(z)$,
$\st\equiv \st(M,z)$, and $\st_0\equiv\st(M_0,z)$, as defined above.
The average overdensity of halos of mass $M$ in spheres of overdensity
$\delta_0$ and radius $R_0$ at epoch $z$, $\delta_h(M,z\vert
R_0,\delta_0)$, can be obtained by dividing number densities given by
eqs. ({\EqEPS}) and ({\EqPS}) and subtracting unity. The halo bias is
then defined as $b\equiv \delta_h/\delta_0$. So far, the bias is
defined in terms of the Lagrangian radius and linearly extrapolated
overdensity. For practical purposes, however, we need the expression
for bias in spheres of observed radius $R$ and (generally non-linear)
overdensity $\delta$.  MW use a spherical collapse model to relate
$(R_0,\delta_0)$ to $(R,\delta)$: $R_0^3=(1+\delta)R^3$ and $\delta_0$
are calculated for given $R$, $R_0$, and $\delta$ using the equations of
spherical collapse. Having made the translation from $(R,\delta)$ to
$(R_0,\delta_0)$, we can calculate the average overdensity of DM halos
in spheres of the observed radius $R$ and overdensity $\delta$ as
\begin{equation}
\delta_h(M,z,z_f\vert R,\delta)=(1+\delta)\frac{n(M,z,z_f\vert R_0,\delta_0)}
{n(M,z,z_f)}-1,
\end{equation}
\begin{equation}
b_{NL}(M,z,z_f,\delta)\equiv \delta_h(M,z,z_f\vert R,\delta)/{\delta},
\end{equation}
where $b_{NL}$ is the halo bias. Note that in general $b_{NL}$ depends
on the overdensity of matter $\delta$ and is therefore non-linear. The
quantities $(R,\delta)$ and $(R_0,\delta_0)$ in eq. ({\Eqdh}) are
related by the spherical collapse model. In the limit of linear
overdensities and large scales, $\delta_0(z)\ll \delta_c(z,z_f)$ and
$M_0\gg M$, the bias is given by
\begin{equation}
b_L(M,z,z_f)=1+\frac{\nu^2-1}{\delta_c(z,z_f)},
\end{equation}
where $\nu\equiv \delta_c(z,z_f)/\sigma(M,z)$. Eq. ({\Eqbl}) shows that
in this regime the bias is linear (does not depend on
$\delta$). Note that the bias of halos with a range of masses
$[M_{min}, M_{max}]$ should be
computed as a mass function weighted average
\begin{equation}
b(z,z_f)= \bar{n}^{-1}(z,z_f)
\int\limits^{M_{max}}_{M_{min}} b(M,z,z_f){\ }n(M,z,z_f){\ }dM ,
\end{equation}
where $\bar{n}(z,z_f)=\int^{M_{max}}_{M_{min}} n(M,z,z_f){\ }dM$
and $n(M,z,z_f)$ is given by eq. ({\EqPS}).

We use eqs. ({\Eqdh})-({\Eqbave}) to calculate the bias predicted by
this analytical model. The linear overdensity $\delta_0$ is calculated
for given $\delta$, $R$, and $R_0=(1+\delta)^{1/3}R$ using the equations of
the spherical collapse model appropriate for our {\LCDM} cosmology
(e.g., Lahav et al. 1991; Eke et al. 1996). The resulting function
$\delta_0(\delta)$ is well described by the fitting formula given by MW
(eq.[18] in their paper), except for $\delta\sim 20-30$, where
deviations reach $\approx 5-10\%$.

\section{Numerical simulation}

We have chosen to study the evolution of the halo bias in a
representative variant of the CDM-type models: the low matter density,
flat, CDM model with cosmological constant (\LCDM):
$\Omega_0=1-\Omega_{\Lambda}=0.3$, $h=0.7$, where $\Omega_0$ and
$\Omega_{\Lambda}$ are present-day matter and vacuum densities, and $h$
is the dimensionless Hubble constant defined as $H_0=100h{\ }{\rm
  km/s/Mpc}$. This model is arguably the most successful model in
matching a variety of existing data.  Observations of galaxy cluster
evolution (Eke et al.  1996), baryon fraction in clusters (Evrard
1997), and high-redshift supernovae (e.g., Perlmutter et al. 1998)
strongly favor the value of matter density $\Omega_0 \approx 0.3$,
while various observational measurements of the Hubble constant (e.g.,
Falco et al. 1997; Salaris \& Cassisi 1998; Madore et al. 1998) tend to
converge on the values of $h\approx 0.6-0.7$. We use a normalization of
the spectrum of fluctuations that is consistent with both observed
cluster abundances (Eke et al. 1996) and the 4-year {\sl COBE} data
(e.g., Bunn \& White 1997): $\sigma_8=1$, where $\sigma_8$ is the rms
fluctuation in spheres of $8h^{-1}{\ }{\rm Mpc}$ comoving radius.
  
The numerical simulation of the {\LCDM} model followed the evolution of
$256^3\approx 1.67\times 10^7$ particles in a periodic $60\hMpc$ box.
The particle mass is thus $\approx 1.1\times 10^9\hMsun$. The
simulation was run using Adaptive Refinement Tree $N$-body code
(ART; Kravtsov, Klypin \& Khokhlov 1997). The ART code reaches high
force resolution by refining all high-density regions with an automated
refinement algorithm.  The refinements are recursive: the refined
regions can also be refined, each subsequent refinement having half of
the previous level's cell size.  This creates an hierarchy of
refinement meshes of different resolution covering regions of interest.
The criterion for refinement is {\em local overdensity} of particles:
in the simulation presented in this paper the code refined an
individual cell only if the density of particles (smoothed with the
cloud-in-cell scheme; Hockney \& Eastwood 1981) was higher than
$n_{th}=5$ particles. Therefore, {\em all} regions with overdensity
higher than $\delta = n_{th}{\ }2^{3L}/\bar{n}$, where $\bar{n}$ is the
average number density of particles in the cube, were refined to the
refinement level $L$. For the simulation presented here, $\bar{n}$ is
$1/8$.  The peak formal dynamic range reached by the code on the
highest refinement level is $32,768$, which corresponds to the smallest
grid cell of $1.83\hkpc$; the actual force resolution is approximately
a factor of two lower (see Kravtsov et al. 1997). The simulation that 
we analyze here has been used in Col\'{\i}n et al. (1998), and we 
refer the reader to this paper for further numerical details.

\section{Halo identification and selection}

Identification of DM halos in the very high-density environments (e.g.,
inside groups and clusters) is a challenging problem. The goal of this
study is to investigate the halo bias at both low and high matter
overdensities, and we therefore need to identify both isolated halos
and satellite halos orbiting within the virial radii of larger systems.
The problems associated with halo identification within high-density
regions are discussed in Klypin et al. (1999; KGKK). In this study we
use a halo finding algorithm called Bound Density Maxima (BDM). The
main idea of the BDM algorithm is to find positions of local maxima in
the density field smoothed at a certain scale and to apply physically
motivated criteria to test whether the identified site corresponds to a
gravitationally bound halo. The detailed description of the algorithm
is given in KGKK and Col\'{\i}n et al. (1998). The publicly available
version of the BDM algorithm used here is described in Klypin \&
Holtzman (1997). The halo catalogs used in the present study were
constructed using numerical parameters given in Col\'{\i}n et al.
(1998).

To construct a halo catalog, we have to define selection criteria based
on particular halo properties.  Halo mass is usually used to define
halo catalogs (e.g., a catalog can be constructed by selecting all
halos in a given mass range).  However, the mass and radius are very
poorly defined for the satellite halos due to tidal stripping which
alters a halo's mass and physical extent (see KGKK). Therefore, we will
use maximum circular velocity $V_{max}$ as a proxy for the halo mass.  This
allows us to avoid complications related to the mass and radius
determination for satellite halos. Moreover, when a halo gets stripped 
$V_{max}$ changes less dramatically than the mass, and is therefore a
better ``label'' of the halo. For isolated halos, $V_{max}$ and the
halo's virial mass are directly related. 

{\footnotesize
\renewcommand{\arraystretch}{1.2}
\renewcommand{\tabcolsep}{1.5mm}
\begin{center}
TABLE 1
\vspace{1mm}

{\sc Halo catalogs}
\vspace{3mm}

\begin{tabular}{cccccc}
\hline
\hline\\
   & 
\multicolumn{2}{c}{$V_{max}>120 {\rm km/s}$} &    &
\multicolumn{2}{c}{$V_{max}>200 {\rm km/s}$}     \\ \\
\cline{2-3} \cline{5-6}\\
z & $M_{vir}^a$ & $N_{halo}$ &   &
$M_{vir}$ & $N_{halo}$\\[3mm]
\hline
      &                     &        &  &                     &\\
$0.0$ & $3.0\times 10^{11}$ & $4707$ &  & $1.4\times 10^{12}$ & $1027$\\
$1.0$ & $2.0\times 10^{11}$ & $7867$ &  & $9.0\times 10^{11}$ & $1443$\\
$2.0$ & $1.1\times 10^{11}$ & $10437$&  & $5.4\times 10^{11}$ & $1675$\\
$3.0$ & $8.0\times 10^{10}$ & $9650$ &  & $3.5\times 10^{11}$ & $1636$\\[2mm]
\hline\\[1mm]
\end{tabular}\\
$^a$ Masses are given in $\hMsun$.
\vspace{3mm}
\end{center}
}
For example, a halo with a
density distribution described by the Navarro, Frenk \& White
(hereafter NFW; 1996, 1997) profile $\rho(r)\propto x^{-1}(1+x)^{-2}$
($x\equiv r/R_s$; $R_s$ is the scale-radius):
\begin{equation}
V_{max}^2=\frac{GM_{vir}}{R_{vir}}\frac{c}{f(c)}\frac{f(2)}{2};
\end{equation}
where $M_{vir}$ and $R_{vir}$ are the virial mass and radius, $f(x)\equiv
\ln(1+x)-x/(1+x)$, $c\equiv R_{vir}/R_s$.  While for the satellite
halos $V_{max}$ may not be related to mass in any obvious way, it is
still the most physically and observationally motivated halo quantity.

We constructed halo catalogs using thresholds in the maximum circular
velocity (i.e. selecting all halos with $V_{max}$ higher than a
specified threshold).  The cluster-size halos are not explicitly
excluded from the halo catalogs. We assume therefore that the center of
each cluster can be associated with a central cluster galaxy. The
latter (due to the lack of hydrodynamics and other relevant processes)
cannot be identified in our simulations in any other way. We limit the
extent of these ``galaxies'' to the central $150\hkpc$ of the cluster.

The redshift dependency of the relationship between halo mass and $V_{max}$ is
expected to evolve because the concentration factor $c$ (see eq.
[\Eqmvmax]) is expected to evolve with redshift. We use the prescription of
NFW to compute the evolution of $c(M,z)$ and to convert $V_{max}$ to the
virial mass, but note that this prescription has been found to deviate
significantly from the results of numerical simulations (Bullock et al.
1998; Eke, Navarro \& Frenk 1998). The values of the virial mass
corresponding to the $V_{max}$ thresholds of $120$ and $200$ km/s used
in our analysis and calculated
using eq. ({\Eqmvmax}) and the NFW prescription for $c(M,z)$ are given
in Table {\TabHaloCat}. This Table also gives the number of halos
at different epochs identified by the halo finder in the simulation box
using these thresholds.

\section{Results}

\begin{figure*}[ht]
\pspicture(0,8.0)(13.0,20.6)
\rput[tl]{0}(0.0,20.8){\epsfysize=9.5cm
\epsffile{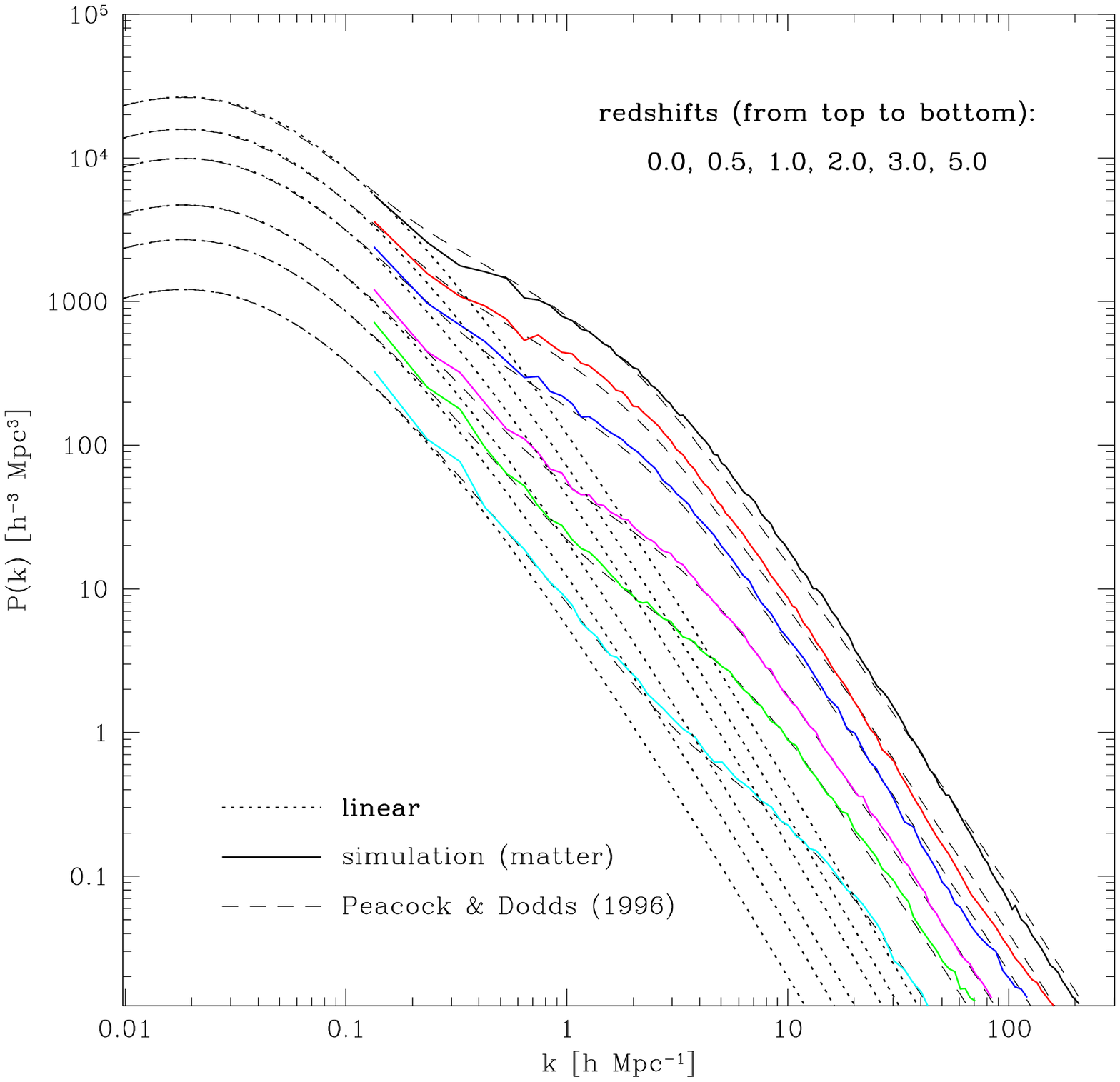}}
\rput[tl]{0}(9.5,20.8){\epsfysize=9.5cm
\epsffile{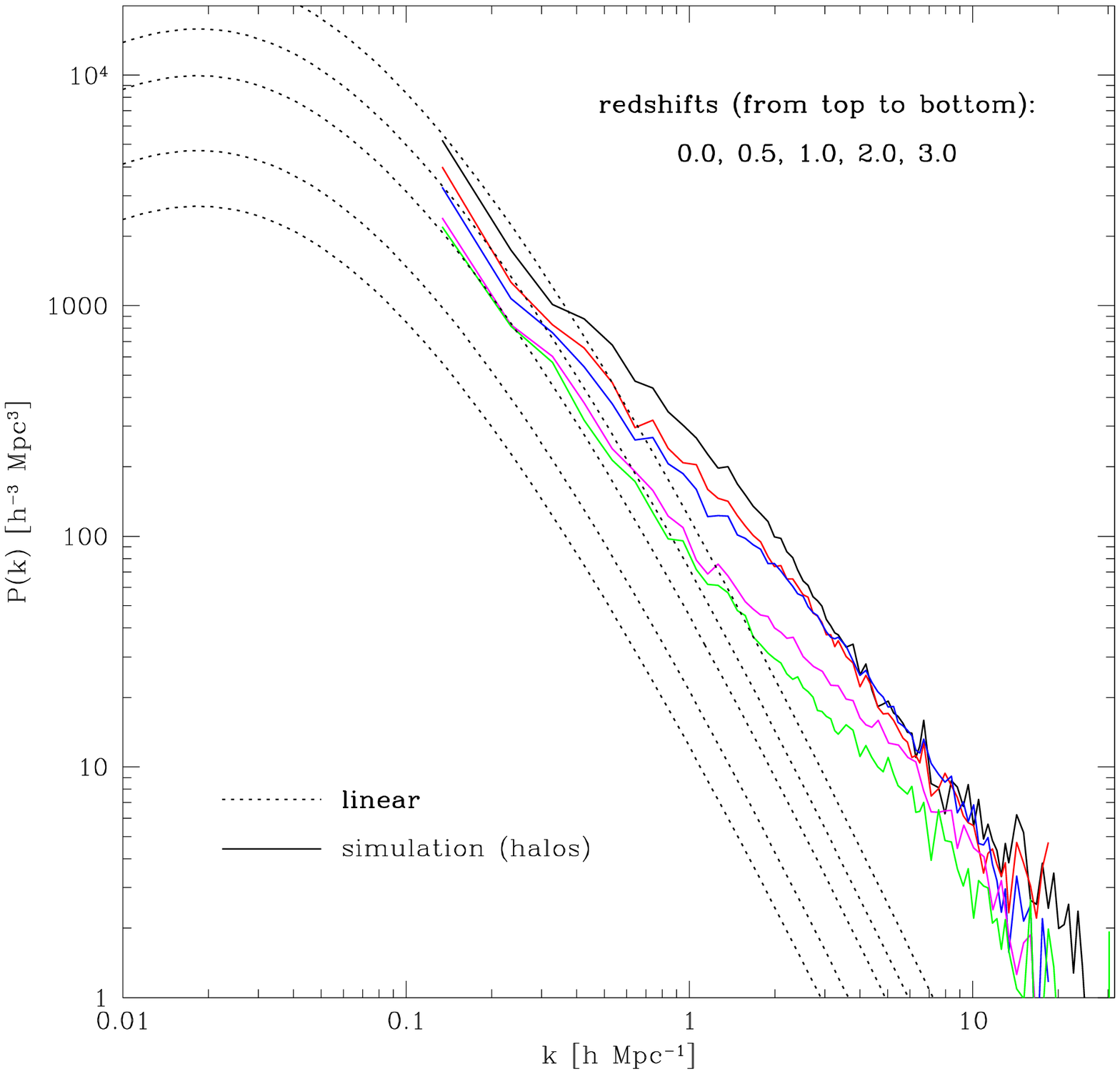}}
\rput[tl]{0}(0.0,11.2){
\begin{minipage}{18.4cm}
  \small\parindent=3.5mm {\sc Fig.}~1.--- Evolution of the (a) matter
  and (b) halo power spectra.  Panel (a): the dark matter power
  spectra, $\pmk$, at different redshifts ({\em solid lines}) are
  compared with the linear spectra ({\em dotted lines}) and predictions
  of the Peacock \& Dodds (1996) fitting formula ({\em dashed lines};
  see text for details). Note that the power spectra in the simulations
  agree with the analytical predictions at all scales, including highly
  non-linear scales at which the ``stable clustering'' approximation
  appears to work well. Panel (b): evolution of the power spectrum for
  halos with maximum circular velocities $>120$ km/s ({\em solid
    lines}) as compared to the linear evolution of the matter power
  spectrum ({\em dotted lines}). Note that the evolution of the halo
  power spectrum, $\phk$, is much slower than that of the matter
  spectrum, $\pmk$, shown in panel (a). At high redshifts the amplitude
  of $\phk$ exceeds that of $\pmk$ by a factor of $\sim 5-10$, while at
  lower redshifts the differences are smaller. The ratio of amplitudes,
  $\phk/\pmk$, depends on the scale. These differences imply that the
  halo bias is time- and scale-dependent.

\end{minipage}
}

\endpspicture
\end{figure*}

\subsection{Evolution of the power spectrum}

Recent numerical and semi-analytical studies have focussed on the bias
evolution as determined from the 2-point correlation function. However,
as new, accurate measurements of the power spectrum at both low
(e.g., Baugh \& Efstathiou 1993, BE93; Gazta\~naga \& Baugh 1998, GB98;
Tadros \& Efstathiou 1996, TE96; Hoyle et al. 1998) and high (Croft et
al. 1998; Weinberg et al. 1998) redshifts become available, it is also
useful to examine and compare the evolution of the power spectra of matter,
$\pmk$, and halo, $\phk$, distributions.

\begin{figure*}[ht]
\pspicture(0,0)(18.5,19.5)

\rput[tl]{0}(1.,19.){\epsfxsize=17cm
\epsffile{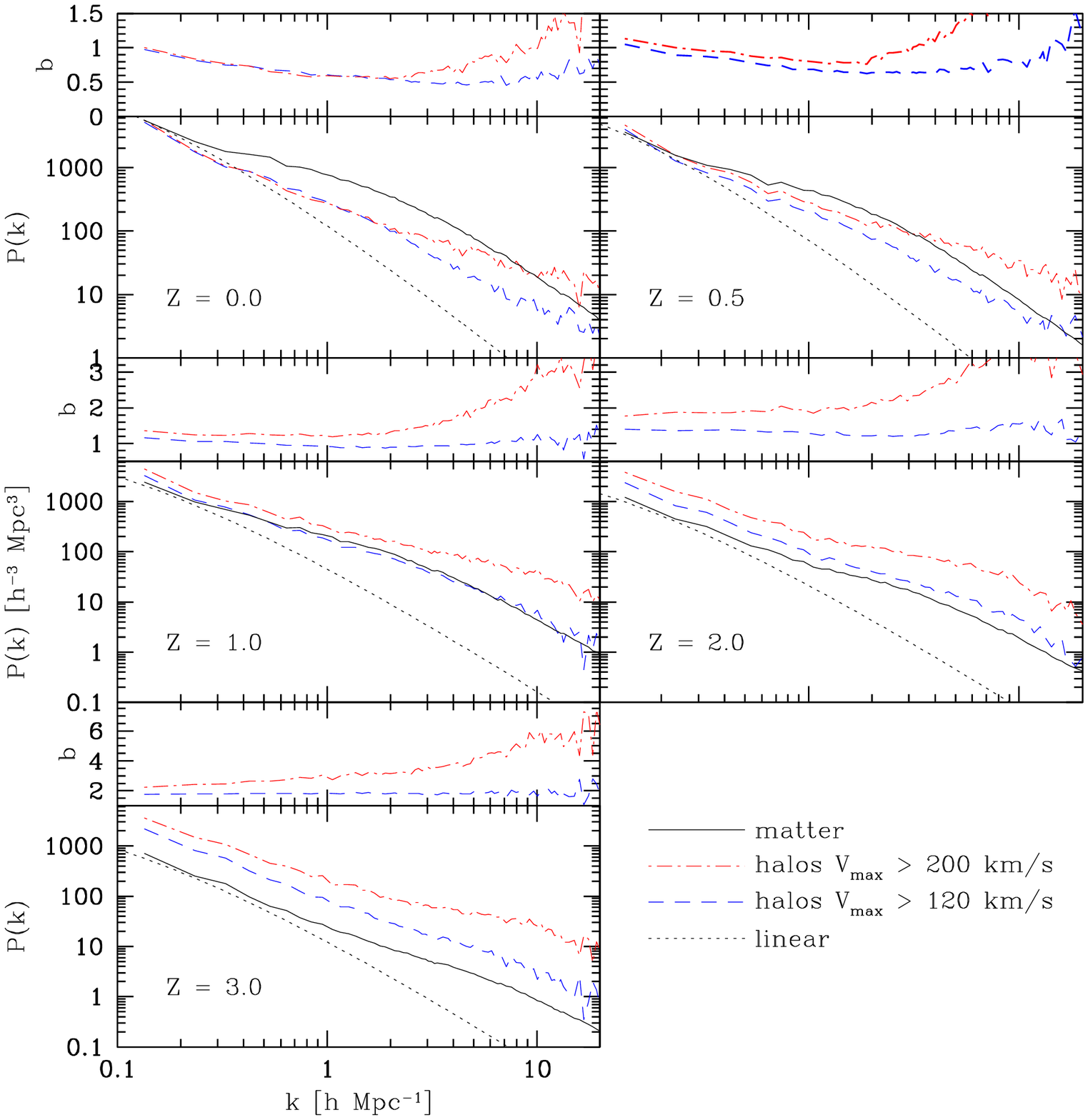}}

\rput[tl]{0}(0.,1.5){
\begin{minipage}{18.5cm}
  \small\parindent=3.5mm {\sc Fig.}~2.--- Evolution of the halo bias
  from $z=3.0$ to the present. The lower portion of each panel shows
  linear ({\em dotted lines}) and non-linear ({\em solid lines})
  spectra of the matter distribution compared to the halo power spectra
  for catalogs with two different lower cutoffs in maximum circular
  velocity: $> 120$ km/s and $>200$ km/s (shown by {\em dashed } and
  {\em dot-dashed} lines, respectively). The upper portion shows the
  corresponding bias $b(k)\equiv [\phk/\pmk]^{1/2}$.
\end{minipage}
}
\endpspicture
\end{figure*}

Figure {\Figpsevol} shows the evolution of {\em real-space} $\pmk$ and
$\phk$ in our simulation. The halo power spectrum is shown for the halo
catalog with the maximum circular velocity threshold of $V_{max}>120{\ 
  }{\rm km/s}$.  To estimate the power spectra over a wide range of
wavenumbers, we have used the method of Jenkins et al. (1998a).  Both $\pmk$
and $\phk$ have been obtained by combining a series of spectra,
$\{P_i(k)\}$, in overlapping ranges of $k$:
$[k_{max}^{i-1},k_{max}^i]$, where $k_{max}^i$ is the maximum
wavenumber at which an accurate estimate of the $P_i(k)$ can be
obtained. To compute $P_i(k)$, the computational cube (of size
$L_{box}$) is divided into $i^3$ subcubes and the particle (or halo)
distributions in these subcubes are superposed.  The FFT of the
resulting density field, estimated using the cloud-in-cell density
assignment scheme (Hockney \& Eastwood 1981), gives an accurate
estimate of the power spectrum in modes that are periodic on scale
the $L_{box}/i$. We have used the FFTs on a $256^3$ grid, and $i=2^m$,
where $m=0,...,6$ and $m=0,...,4$ for the dark matter and halos,
respectively. Comparisons with direct $512^3$-grid FFT spectra
suggested the use of $k_{max}^i=k_{Ny}^i/6$, where
$k_{Ny}^i=128i(2\pi/L_{box})$ is the Nyquist wavenumber for $P_i(k)$
($k_{max}^0=2\pi/L_{box}$).  The individual power spectra have been
corrected for the shot noise by subtracting the noise spectrum
estimated using the same method.

Panel (a) of figure {\Figpsevol} shows the evolution of the matter power
spectrum from $z=5$ to the present. For comparison we also show the
non-linear evolution predicted by the fitting formula of Peacock \&
Dodds (1996, hereafter PD96):
$\Delta^2_{NL}(k_{NL},z)=f_{NL}[\Delta^2_L(k_L)]$, where
$\Delta^2(k)=dP(k,z)/d\ln k$ and subscripts $L$ and $NL$ denote linear
and non-linear quantities, respectively. The analytical expression for
$f_{NL}$ depends on five fitting parameters (see \S 3.3 in PD96)
obtained by fitting the power spectra of the scale-free $N$-body
simulations.  Although there are some minor difficulties in applying
these fitting results to the realistic models with scale-dependent
power-law spectral index (PD96; Smith et al. 1998; Jenkins et al.
1998a), the prediction works remarkably well. We have been able to match
the non-linear spectra in our simulation at all epochs with only small
change in the fitting parameters of PD96.  Namely, we have used
$n_{eff}(k_L)=d\ln P(k)/d\ln k\vert_{k_L}$ (as opposed to an
alternative $n_{eff}(k_L/2)$) for the estimate of the spectral index at
wavenumber $k_L$ and slightly changed the power-index dependence for the
fitting parameter $V$. This parameter controls the amplitude of the high-$k$
asymptote; instead of using $V=11.55(1+n_{eff}/3)^{-\eta}$, with a
single time-independent value of $\eta=0.423$ given by PD96, we use
$\eta$ varying from $0.7$ at $z=5$ to $0.45$ at $z=0$. The PD96
formula, with their fitting parameters unchanged, fares well at $z<1$,
but underpredicts the amplitude of the asymptote at high $z$. Figure
{\Figpsevola} shows that, with this small modification\footnote{This
  modification is, of course, arbitrary and the same result could 
  possibly be achieved with other changes; for example, by varying
  the wavenumber at which $n_{eff}$ is computed.}, the PD96 prediction is a
success. Nevertheless, if the desired accuracy of the non-linear
spectrum estimate is $\lesssim 50\%$, the necessity of such (generally
time- and cosmology-dependent) modifications should be kept in mind.

Figure {\Figpsevolb} shows that the evolution of $\phk$ is much slower than
the evolution of the matter power spectrum. Although the scale of
non-linearity (wavenumber of an upward inflection of $\phk$) is seen
clearly in the halo spectra and approximately matches the corresponding
scale in $\pmk$ at the same epoch, the shapes of the halo and matter
power spectra are quite different: the $\phk$, for example, can be
approximated well by simple power-law, while the shape of $\pmk$ is
more complicated. This difference, together with the difference in the
evolution rate, means that the bias of the halo distribution is time- and
scale-dependent. A similar conclusion has been reached by many
researchers from comparisons of halo and matter two-point correlation
functions (e.g., recently, Bagla 1998; Col\'{\i}n et al. 1998; Kauffman
et al. 1998ab; Katz, Hernquist \& Weinberg 1998; and references therein).
Note, however, that as we mentioned in \S 1, the scale-dependence of the
bias is different in real- and $k$-space (see eq.[\EqbPbxi]). Although
we will be referring to both $b_P(k)$ and $b_{\xi}(r)$ (defined in eqs.
[\Eqbxi] and [\EqbP]) as the bias functions, it should be kept in mind
that they have different functional forms.

Figure {\Figpsb5p} shows the evolution of the halo bias, $b(k)\equiv
[\phk/\pmk]^{1/2}$, from $z=3.0$ to the present epoch for two halo
catalogs selected with low and high-mass thresholds: $V_{max}>120{\ 
  }{\rm km/s}$ and $V_{max}>200{\ }{\rm km/s}$. The bias evolves from a
value of $\approx 2-4$ at $z=3$ to $\approx 1$ at $z=0$. At high
redshifts, the bias depends on the selection threshold indicating that
it is mass-dependent; the differences, however, become progressively
smaller for lower redshifts. Also, the scale-dependence of the bias of
the lower-mass halos, albeit being redshift-dependent and generally
$\neq 1$, is significantly weaker than the scale-dependence of the
higher-mass halos. At $z=1$, for instance, the power spectrum of
$V_{max}>120{\ }{\rm km/s}$ halos follows that of the mass almost
exactly at all probed wavenumbers. In agreement with analytical
prediction of Scherrer \& Weinberg (1998), the bias is virtually
scale-independent at linear scales $k<k_{nl}$, where $k_{nl}$ is the
wavenumber where the $\pmk$ becomes non-linear and exceeds the linear
prediction ($k_{nl}\approx 0.2-0.4$). Note also that during the
evolution the bias at these linear scales is driven to the value of
$\approx 1$, as expected in the linear regime (Tegmark \& Peebles
1998). This is true for both low- and high-mass catalogs, which
indicates that the bias evolution at these scales is driven by
gravitational growth of clustering that tends to erase any initial
(mass-dependent) differences in the halo and mass distributions
(Tegmark \& Peebles 1998).

The bias evolution at non-linear ($k>k_{nl}$) scales is more
complicated.  The bias evolves to values less than unity at $z<1$, and
its scale-dependence becomes progressively stronger over a wider range of
wavenumbers. We will discuss the possible interpretation of the bias
evolution in the non-linear regime in the following section. However,
we would like to point out here that the net result of this evolution
is the $\phk$ at $z=0$ which is {\em significantly anti-biased} with
respect the overall matter distribution but which agrees very well with
the power spectrum of observed galaxy distribution. Figure {\Figpsz0}
compares the $z=0$ power spectra of halos and matter in our simulation
with the power spectrum of galaxies (BE93; GB98; TE96) in the APM
survey. At small scales ($k\gtrsim 2h {\rm Mpc^{-1}}$), the $\phk$
depends on the catalog's $V_{max}$ threshold. The galaxy power
spectrum, however, lies comfortably in between two likely possibilities
of galaxy mass cutoffs. The maximum circular velocities of $120$ and
$200$ km/s correspond at $z=0$ to the virial masses of $\approx 3\times
10^{11}\hMsun$ and $\approx 1.5\times 10^{12}\hMsun$, respectively. At
larger scales, the power spectra of halos and galaxies agree within the
errors. The ``inflection'' in the galaxy power spectrum (GB98) is
reproduced at the correct wavenumber of $k\approx 0.2h{\ }{\rm
  Mpc^{-1}}$.
 {\pspicture(0.5,-1.5)(13.0,14.5)
\rput[tl]{0}(0.2,14.7){\epsfxsize=9.7cm
\epsffile{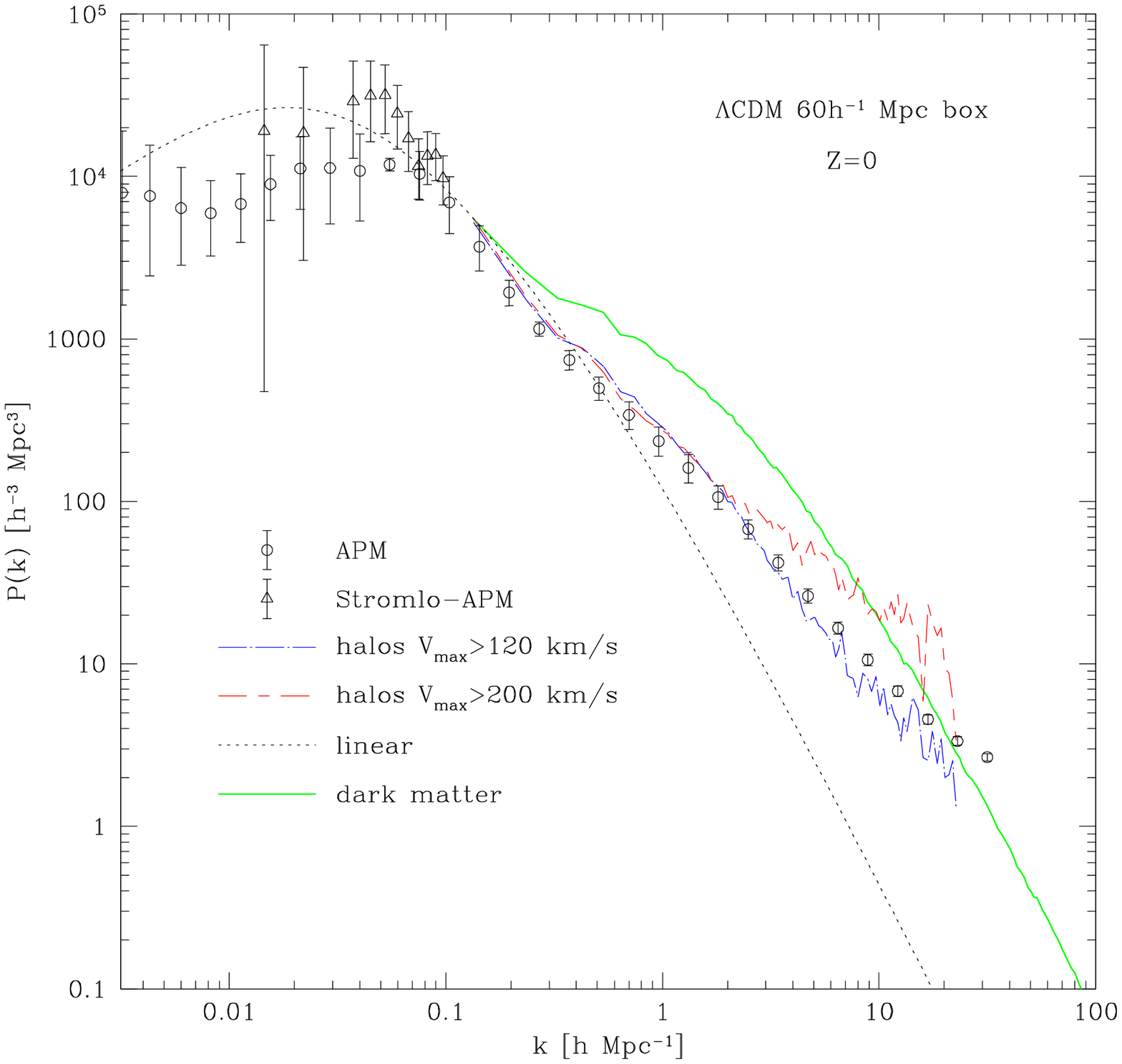}}
\rput[tl]{0}(0.5,4.8){
\begin{minipage}{8.7cm}
  \small\parindent=3.5mm {\sc Fig.}~3.--- Comparison of the $z=0$ power
  spectra of the matter ({\em solid line}) and halo ({\em dashed} and
  {\em dot-dashed} lines) distributions with the galaxy power spectra
  estimated using the APM (Gazta\~naga \& Baugh 1998) and Stromlo-APM
  surveys (Tadros \& Efstathiou 1996); the latter is measured in {\em
    redshift space} and must be reduced by $\approx 40\%$ to correct
  out the redshift-space effects (see \S {\SecPSEvol} for details). The
  errorbars of the galaxy power spectra are $2\sigma$.  For clarity,
  the Stromlo-APM power spectrum is shown only at $k<0.1h{\ }{\rm
    Mpc^{-1}}$. Note that the halo power spectrum, $\phk$, matches the
  galaxy power spectrum significantly better than does the spectrum of
  the overall matter distribution, $\pmk$. Both galaxy and halo
  distributions are {\em anti-biased} at $k\sim 0.2-10{\ }{\rm
    Mpc^{-1}}$; the $\phk$ at these scales has a shape and amplitude
  different from those of $\pmk$. At larger scales, $k\lesssim 0.2{\ 
    }{\rm Mpc^{-1}}$, the halo distribution is approximately unbiased.
\end{minipage}
 }
\endpspicture}
 We interpret the ``inflection'' wavenumber in the
observed spectrum as the scale of non-linearity, because both halo and
matter power spectra in the simulation become non-linear at higher $k$.
The amplitude of the matter power spectrum at these $k$, however, is
$\sim 3-4$ times higher and does not come anywhere close to matching
the observed power spectrum. This shows clearly that it is crucial to
consider the distribution of halos, not matter, when comparing model
predictions to the observations.

It is not clear whether the power spectrum in the {\LCDM} model
considered here can reproduce the amplitude and shape of the turnover
observed in the galaxy power spectrum at $k\sim 0.01-0.06h{\ }{\rm
  Mpc^{-1}}$. Figure {\Figpsz0} shows that the {\LCDM} spectrum does
not reproduce the turnover in the spectrum recovered from the APM
angular correlation function (BE93, GB98); however, it is in better
agreement with the power spectrum derived by TE96 from the Stromlo-APM
redshift survey. Also, recently published power spectrum of the
Durham/UKST survey (Hoyle et al. 1998) agrees very well with the
Stromlo-APM spectrum and with the spectrum of the {\LCDM} model.  The
Stromlo-APM and Durham/UKST spectra have been computed in {\em
  redshift-space}, but the differences from the real-space spectrum
{\em in the} {\LCDM} {\em model} at these scales are expected to be
$\lesssim 40\%$ (TE96; Smith et al.  1998). Even with the $40\%$ lower
amplitude, the spectra are in agreement with the model and, within
$2\sigma$ errors, with the APM power spectrum.  The latter indicates
either that there are systematic differences between the galaxy surveys
or that the cosmological model is incorrect (because it predicts an
incorrect redshift-to-real space correction). There is also a
possibility that the amplitude and errors of the APM power spectrum are
somewhat underestimated at these large scales (Peacock 1997).
Unfortunately, at present, statistical and systematic observational
errors are too large at these scales to be able to make a decisive
conclusion. In any case, considering the whole range of observationally
probed wavenumbers, the agreement between the data and the model is
much better when we compare observations to $\phk$, as opposed to
$\pmk$.

\begin{figure*}[ht]
\pspicture(0,0)(18.5,19.5)

\rput[tl]{0}(1.,20.5){\epsfxsize=17cm
\epsffile{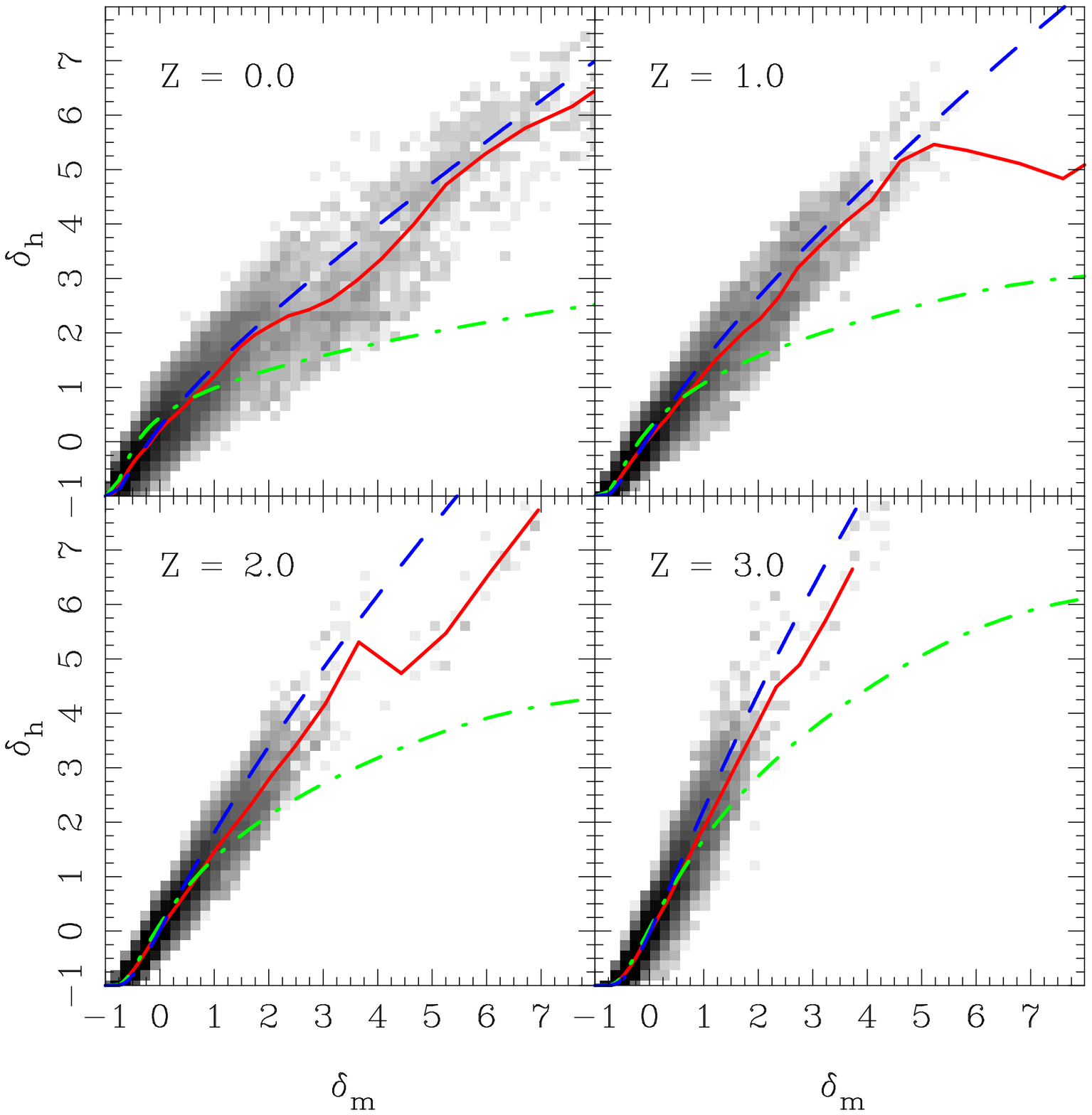}}

\rput[tl]{0}(0.,4.){
\begin{minipage}{18.5cm}
  \small\parindent=3.5mm {\sc Fig.}~4.--- The overdensity of halos
  $\delta_h$ vs. the overdensity of matter $\delta_m$ at different
  epochs (indicated on each panel). Both overdensities has been
  estimated in spheres of radius $R_{\rm TH}=5\hMpc$ randomly placed in
  the simulation box. The solid curves show the {\em average} relation,
  calculated by averaging $\delta_h$ of the individual spheres in bins
  of $\delta_m$.  The actual binned distribution of the $\delta_m$ and
  $\delta_h$ values is shown by the shades of grey on a 2D grid, where
  the intensity of grey corresponds to the natural logarithm of the
  number of spheres in this grid cell.  The {\em long-dashed} and {\em
    dot-dashed} curves show predictions of the analytical model
  described in \S {\Secbanalytic}. The dot-dashed curve is a prediction
  of the model assuming that formation redshift coincides with
  observation redshift ($z_f=z$), while the long-dashed curve
  corresponds to assumption $z_f=z+1$ (see \S {\Secddl} for details).
  The figure shows that at all epochs halo overdensity is tightly
  correlated with overdensity of matter. However, the slope of the
  correlation, the bias, depends on $\delta_m$ (i.e., the bias is
  non-linear) and changes non-trivially with time. Note that at $z>1$
  the halo distribution is biased in overdense regions
  ($\delta_m\gtrsim 0$) but is anti-biased in underdense regions
  ($\delta_m\lesssim -0.5$). At low-redshift there is an anti-bias at
  high-overdensities and almost no bias at low overdensities (see \S
  {\SecDiscussion} for discussion).
\end{minipage}
}
\endpspicture
\end{figure*}

\subsection{Overdensity of matter vs. overdensity of halos}

\subsubsection{Linear and mildly non-linear overdensities}

The evolution of bias, as determined from the power spectra in the
previous section, agrees qualitatively with the bias evolution derived
from the correlation function analyses (e.g., recently, Bagla 1998;
Col\'{\i}n et al. 1998; Kauffman et al. 1998ab; Katz, Hernquist \&
Weinberg 1998; and references therein). The bias functions
$b_P(k)\equiv \sqrt{P_h(k)/P_m(k)}$ and $b_{\xi}(r) \equiv
\sqrt{\xi_h(r)/\xi_m(r)}$ (see \S {\SecPSEvol}), defined using the
power spectrum and the correlation function, illustrate
the scale dependence of the bias. However, it is difficult to interpret
$b_P$ and $b_{\xi}$ in terms of the most generic definition of
bias\footnote{All definitions are, of course, equivalent if the bias is
linear. This, however, is not true for the non-linear
  bias, which appears to be a generic feature of the CDM models. }:
$\delta_h=b_{\delta}\delta_m$, where we denote the bias in this
definition by $b_{\delta}$ to distinguish it explicitly from $b_P$ and
$b_{\xi}$. The $b_{\delta}$ shows how bias depends on the matter
density at a fixed scale, the information which cannot be extracted
easily from $b_{\xi}$ and $b_P$. Therefore, to get a full picture of
the bias evolution, we examine the evolution of $b_{\delta}$ in our
simulation.

To estimate $\delta_h$ and $\delta_m$, we use the top-hat filter (see
\S {\Secbanalytic}) of {\em comoving} radius $R=5\hMpc$. The size of
the radius is a compromise between halo statistics and the range of 
probed overdensities. Note that our simulation box contains only $216$
independent spheres of this size. This is the maximum number of spheres
that can be used to study the scatter of the bias. However, we are
primarily interested in the average behavior of $b_{\delta}$;
therefore, in order to probe the wide range of overdensities, we use a
large number ($50,000$) of spheres, placed randomly throughout the
simulation box. To calculate $\delta_h$, we have used halos with
maximum circular velocities of $V_{max}>120\kms$. Due to limited mass
resolution, the halo catalogs are somewhat incomplete for
$V_{max}\lesssim 100\kms$ (see Gottl\"ober, Klypin \& Kravtsov 1998).
The catalog with the threshold of $120\kms$ is complete and contains a
large enough number of halos (see Table {\TabHaloCat}) to provide
sufficient statistics.

Figure {\Figb4plin} shows overdensity of dark matter halos, $\delta_h$,
as a function of matter overdensity $\delta_m$ at four epochs: $z=0, 1,
2, 3$. The solid curves show the {\em average} relation, calculated by
averaging $\delta_h$ of the individual spheres in bins of $\delta_m$. 
The actual binned distribution of the $\delta_m$ and $\delta_h$ values is
shown by grey-scale shades on a grid, where the intensity of grey
corresponds to the natural logarithm of the density of points in the
grid cell. The figure shows that at all epochs, the halo overdensity is
tightly correlated with the overdensity of matter. However, the slope of
the correlation, the bias, depends on $\delta_m$ (i.e., the bias is
non-linear) and changes non-trivially with time. 

The dashed and dot-dashed lines in each panel of Figure {\Figb4plin}
show predictions of the analytical model of bias described in \S
{\Secbanalytic} (namely, eqs. [{\Eqdh}] and [{\Eqbnl}]). To account for
the range of halo masses used in our halo catalog, we calculate the
mass function weighted bias (eq. [\Eqbave]) using
$M_{max}=10^{13}\hMsun$ as an upper limit of integration, and
redshift-dependent $M_{min}(z)$ corresponding to our selection
threshold of maximum circular velocity of $120\kms$ (see \S
{\SecSimulation} and Table {\TabHaloCat}).

The two model curves correspond to different assumptions about
formation redshift of halos, $z_f$: $z_f=z$ (i.e., halos forming at the
epoch of observation) for the dot-dashed curve and $z_f=z+1$ for the
dashed curve. In the standard Press-Schechter model, hierarchical
build-up of halos is a continuous process and, therefore, if mass is
considered as a defining property of a halo, $z_f=z$.  However, this
interpretation fails if halos can retain their identity over a finite
period of time (e.g., see discussion in Moscardini et al.  1998).  For
example, if a halo is accreted by a more massive system and orbits in
the system's potential instead of merging instantly, it can be
identified at $z<z_f$. The galaxy-conserving model of bias evolution
(e.g., Moscardini et al. 1998 and references therein) represents an
extreme in which halos are never destroyed after being formed and in
which the halo clustering is driven solely by the gravitational pull
from the surrounding growing structures.

In reality, we expect the merging to take place and the halos to have
individual merging histories and thus individual {\em survival
  times}\footnote{The survival time is the time between halo formation
  and destruction.}  (Lacey \& Cole 1993; Kitayama \& Suto 1996).  A
rigorous treatment of bias should therefore take into account only
the halos that survive until the epoch of observation
$z$. In practice, this is a difficult task: although the halo formation
epoch is well defined, definition of the halo destruction is not
trivial. Lacey \& Cole (1993) define halo lifetime as the period between
formation time of a halo and the time by which this halo is
incorporated into a more massive system. This definition differs
significantly from how we define the destruction when analyzing the
simulations: the halo is destroyed either when it merges with another
halo and looses its identity or when the tidal stripping brings the
mass bound to the halo below some mass threshold (defined as a
selection criterion or by the mass resolution of the simulation).
Therefore, for illustration purposes, we will treat $z_f$ as a free
parameter, leaving a more rigorous treatment for future work.

Figure {\Figb4plin} shows that different assumptions about $z_f$ lead
to significant difference in the behavior of bias predicted by eq.
({\Eqbnl}). Although at $\delta_m\lesssim 0.5$ the two bias predictions
are similar, at higher overdensities they diverge,the $z_f=z+1$ assumption
providing a much better match to the simulation results. Note that
$z_f=z$ underestimates the bias in the simulation at
$\delta_m\gtrsim 1$.  This was noted recently by Jing (1998) who
studied mass- and scale-dependence of bias in the linear regime. For the
simulation of the {\LCDM} model used in this paper, Jing finds that
halos of mass $10^{11}\hMsun$ exhibit bias of
$b^2(M=10^{11}\hMsun)\approx 0.5-0.6$, while {\em linear} MW bias is
(eq. [\Eqbl] assuming $z=z_f$): $b_L^2\approx 0.28$. For the same halo
mass and $z_f=z+1=1$, eq. (\Eqbl) gives $b_L^2\approx 0.67$, and
therefore the finite survival time of halos may naturally explain the
bias discrepancy. It is worth noting that for cluster halos, the
$b_L$ with $z_f=z$ provides a considerably better approximation
which likely reflects late cluster formation ($z\approx z_f$)
and/or smaller survival times for cluster-size halos, as is indeed
predicted by the extended Press-Schechter theory (Lacey \& Cole 1993). 

\subsubsection{Non-linear overdensities}

Figure {\Figbnonlin} shows the present-day $\delta_h$-$\delta_m$
relation at higher overdensities. Although we use a large number of
spheres 
 {\pspicture(0.5,-1.5)(13.0,13.5)
\rput[tl]{0}(0.,13.7){\epsfxsize=9.7cm
\epsffile{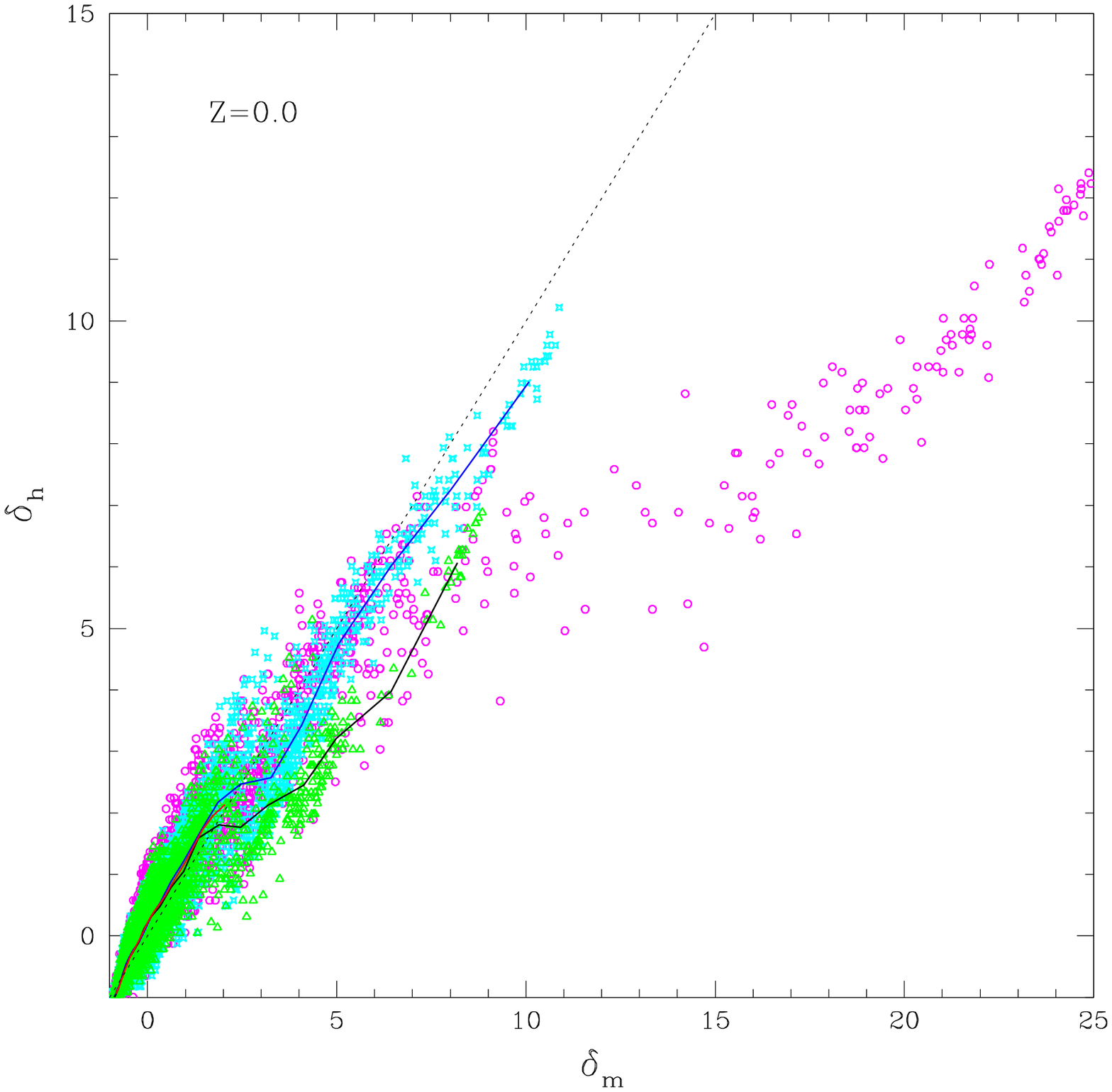}}
\rput[tl]{0}(0.5,3.8){
\begin{minipage}{8.7cm}
  \small\parindent=3.5mm {\sc Fig.}~5.--- The overdensity of halos
  $\delta_h$ vs. the overdensity of matter $\delta_m$ at $z=0$ in
  spheres of radius $R_{\rm TH}=5\hMpc$ randomly placed in the
  simulation box. The different symbols correspond to the spheres in
  three independent (non-overlapping) subcubes of the simulation volume
  (sub-cube size of $30\hMpc$) that contain most of the massive
  clusters in the simulation. The lines represent the average relations
  for the sub-cubes marked with triangles and crosses.  Notice the
  differences in $\delta_h$ vs.  $\delta_m$ correlation at high
  $\delta_m$. We attribute these differences to the differences in
  formation histories of structure in these sub-cubes (see \S
  {\SecDiscussion} for details). The anti-bias at very high
  overdensities ($\delta_m\gtrsim 10$) arises in a region surrounding
  the richest Coma-size cluster in the simulation.
\end{minipage}
 }
\endpspicture}
and oversample the density field, the spheres in independent
regions of space are independent and may thus give an idea about the
true scatter in the halo bias. In order to study this scatter, we have
divided the computational box into eight equal-size ($30\hMpc$)
non-overlapping sub-cubes. Different symbols (of different colors) in
Figure {\Figbnonlin} correspond to spheres in the three sub-cubes that
contain most of the massive clusters in the simulation at $z=0$. The
other five sub-cubes contain lower density regions and are not shown
for clarity.

Figure {\Figbnonlin} shows that the differences between
$\delta_h$-$\delta_m$ relations in different sub-cubes are
significantly larger than the scatter of each individual relation.
Note, however, that even within a single sub-cube there are
indications of real scatter: the sub-cube represented by the triangle
markers shows a dichotomy of $\delta_h$ at fixed $\delta_h\approx 2-4$.
Analysis of the sub-cubes has shown that differences in the
$\delta_h$-$\delta_m$ correlation among the sub-cubes are real and are
caused by the differences in the non-linear structures they contain.

It appears that the scatter and the differences can be explained by the
following two effects.  The centers of the spheres with
$\delta_m\gtrsim 5$ fall preferentially in close vicinity to the
massive group- and cluster-size halos. These cluster halos span a wide
range of masses and, most importantly, formation times. The differences
in formation times result in different fates of dark matter halos
orbiting inside these clusters. If, for example, a cluster forms and
accretes the bulk of its mass and halos early, the halos have time to
suffer substantial losses of mass due to tidal stripping and losses
of orbital energy from dynamical friction. Dynamical friction
may lead to a merging between satellite and central cluster object,
thus resulting in a ``loss'' of the satellite. Tidal stripping may
lead to a substantial mass loss and may ultimately (although at a much
slower rate) decrease the halo's maximum circular velocity (see KGKK)
bringing the halo below our threshold $V_{max}$.  These effects are
enhanced because the virial radius is smaller at earlier epochs and so
are the typical pericenters of satellite orbits.  If, on the other
hand, the cluster forms late and accretes most of its mass fast and at
relatively low redshifts, the halos are accreted onto a massive cluster
and thus have higher orbital energies and orbits with larger
pericenters. Moreover, many halos simply do not have enough time to
suffer substantial tidal or orbital energy losses.

The differences in formation histories may therefore lead to
significant differences in the halo content among clusters. This is
illustrated in Figure {\Figbpro} which shows ``bias profiles'': the
ratios of the {\em integral} overdensities of halos and matter in
spheres of increasing radii centered on a cluster. The profiles for 5
of the clusters from the three sub-cubes of Figure {\Figbnonlin} are
shown at three different epochs $z=0.0,0.5,1.0$. Note the large 
differences between profiles at $z=1$. While the cluster marked CL1
exhibits strong {\em anti-bias} (i.e., low $\delta_h/\delta_m$), the
halos in cluster CL2 are strongly biased. Similar differences are seen
in the rest of clusters. Interesting differences can also be observed 
in the subsequent evolution of $b(r)$. Cluster CL1 shows very mild
evolution in $b(r)$ at small ($r\lesssim 1\hMpc$) scales, whereas 
other clusters show very strong evolution between $z=1.0$ and $z=0.5$,
and much weaker evolution from $z=0.5$ until present. 

The changes in the rate of evolution may seem counterintuitive; indeed, the
time period corresponding to the $z=1.0-0.5$ interval is approximately
twice as short ($\approx 2.5$ Gyrs) as the period between $z=0.5$ and
$z=0$ ($\approx 5$ Gyrs). We could thus expect more significant changes
at $z<0.5$ due to more prolonged effects of tidal stripping and dynamical
friction. However, as we have noted above, as the mass of a cluster
grows with time, the halos are accreted on higher-pericenter,
higher-energy orbits and thus are not as likely to approach the dense
central region or spend a substantial amount of time there.

\begin{figure*}[ht]
\pspicture(0,0)(18.5,18.)

\rput[tl]{0}(1.,19.){\epsfxsize=17cm
\epsffile{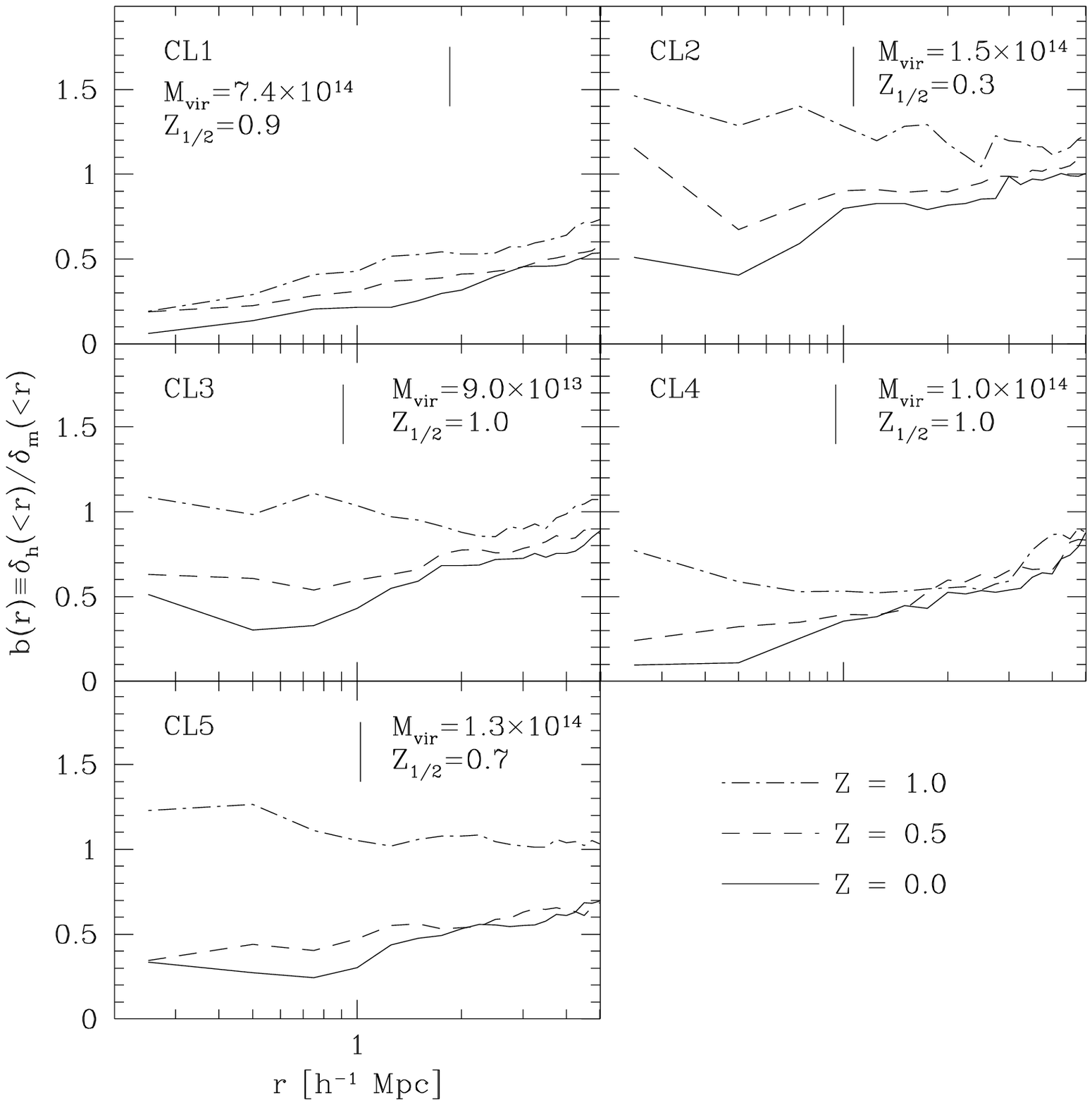}}

\rput[tl]{0}(0.,2.75){
\begin{minipage}{18.5cm}
  \small\parindent=3.5mm {\sc Fig.}~6.--- Bias profiles (ratio of halo
  to matter overdensities inside a sphere of radius $r$ centered on the
  cluster center) of five of the rich ($\Mvir\gtrsim 10^{14}\hMsun$)
  clusters from the three sub-cubes shown in Fig. {\Figbnonlin} at
  three different epochs $z=0.0,0.5,1.0$. The vertical lines indicate
  the virial radii of clusters at $z=0$; $\Mvir$ and $Z_{1/2}$ indicate
  the $z=0$ virial mass and redshift at which cluster had half of this
  mass, respectively. Note the large differences between profiles at
  $z=1$.  While cluster marked CL1, the most massive cluster in the
  simulation, exhibits strong {\em anti-bias} (i.e., low
  $\delta_h/\delta_m$), the halos in cluster CL2 are strongly biased.
  Similar differences are seen in the rest of clusters. Interesting
  differences can also be observed in the subsequent evolution of
  $b(r)$. Cluster CL1 shows very mild evolution in $b(r)$ at small
  ($r\lesssim 1\hMpc$) scales, whereas other clusters show very strong
  evolution between $z=1$ and $z=0.5$, and much weaker evolution from
  $z=0.5$ until present.
\end{minipage}
}
\endpspicture
\end{figure*}

To illustrate that this indeed is the case, we have analyzed the dynamical
evolution of halos identified within the virial radius of clusters at
different moments in time. Figure {\FigclevolI} shows the evolution of 5
halos randomly selected within the virial radius of cluster CL1 at $z=3$
(there are a total of 10 halos within the virial radius). At all epochs, 
CL1 is the most massive cluster in the simulation box; at $z=3$ its mass
was $1.5\times 10^{13}\hMsun$. The panels in each of the horizontal
rows in Figure {\FigclevolI}, show the evolution of particles that bound to
the halos at $z=3$. The orbits of 4 halos are contained within the cluster
virial radius, $\approx 150\hkpc$, at $z=3$ (shown as a circle in each
panel), where all of these halos merge with the central $\sim 100\hkpc$
size object by $z=1$ (i.e., after $\approx 3.5$ Gyrs). The halo on the
most eccentric orbit (bottom row) survives until $z\approx 0.5$ and
gets tidally destroyed after that. 

For comparison, Figure {\FigclevolII} shows the evolution of 10 halos
randomly selected within the virial radius of the same cluster CL1 at
$z=0.5$. As before, most of the halos stay within the $z=0.5$ virial
radius ($\approx 1.2\hMpc$). However, unlike the $z=3$ halos, most of
them (8 out of 10) survive until $z=0$ (i.e., during the period of
$\approx 5$ Gyrs). Although some halos suffer substantial mass loss in
their outer regions, the dense halo cores can be identified at $z=0$. 
Note that two halos on low-pericenter orbits do merge with the central 
object. 

Finally, we have also visually examined the fate of the halos identified in
cluster CL5 of Fig. {\Figbpro} at $z=1$ and $z=0.5$. More than half
of the $z=1$ halos merge with the central object by $z=0.5$, while
most of the $z=0.5$ halos survive until $z=0$.  The difference is
caused by both the lower typical orbit pericenters at higher redshifts
and the higher efficiency of the dynamical friction due to a smaller
mass of cluster. It explains the evolution of the bias profile shown in
Fig. {\Figbpro}.

\begin{figure*}[ht]
\pspicture(0,0)(18.5,18.)

\rput[tl]{0}(1.,17.5){\epsfxsize=17cm
\epsffile{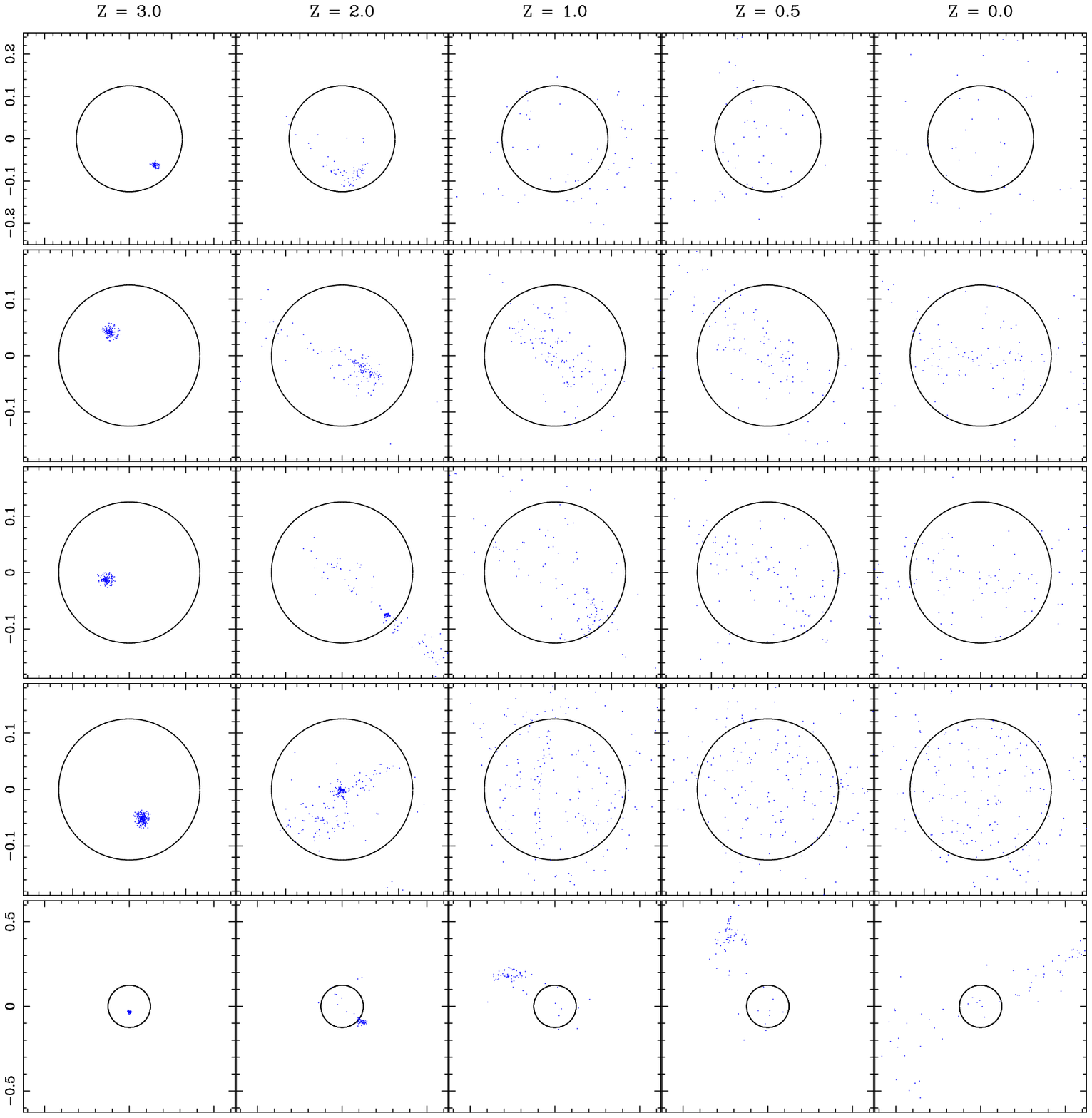}}

\rput[tl]{0}(0.,1.5){
\begin{minipage}{18.5cm}
  \small\parindent=3.5mm {\sc Fig.}~7.--- The leftmost column of panels
  shows particles bound to the five halos identified within the virial
  radius of the cluster CL1 shown in Fig. {\Figbpro} at $z=3$
  ($\Rvir\approx 120\hkpc$ (proper) and $\Mvir\approx 1.5\times
  10^{13}\hMsun$ at $z=3$). The rest of the columns show positions of
  the same particles at later moments. Four out of five halos merge
  with the central cluster halo by $z\approx 1$. In all panels
  particles are plotted in proper coordinates; the circles in all
  panels have the same radius equal to the virial radius of the cluster
  at $z=3$.
\end{minipage}
}
\endpspicture
\end{figure*}

To illistrate the relative efficiency of dynamical friction at
different epochs of cluster evolution, we present the evolution of the
dynamical friction time-scale in a cluster. At a given epoch, the
dynamical friction time, $t_{fric}$, can be estimated using
Chandrasekhar's formula (Binney \& Tremaine 1987), assuming the cluster
mass and density distribution and the mass of the satellite. In the
right column of Figure {\Figzmerge}, we present estimates of $t_{fric}$
for halos with maximum circular velocity $V_{max}=120{\ }{\rm km/s}$
for clusters of different final masses. The mass accretion histories of
clusters of similar present-day mass exhibit a spread around an
average, typical for this mass, evolution track. To account for this
spread, we have used both the average mass evolution tracks and two
individual evolution tracks representative of the early- and
late-forming tails of the population (these tracks represent $\approx
2\sigma$ deviations from the average mass evolution track). The cluster
mass evolution tracks used here (Avila-Reese \& Firmani 1997) have been
generated using the Monte-Carlo method of Lacey \& Cole (1993). For each
epoch, we compute $t_{fric}$ using eqs. (8-10) of KGKK assuming the NFW
density distribution (with an appropriate $c(M,z)$, see \S {\SecHalo})
for both the cluster and satellite at a distance $R_{vir}/2$ from the
cluster center ( where $R_{vir}$ is the virial radius of the cluster at
this epoch), and accounting explicitly for the mass loss due to the
tidal stripping. 

\begin{figure*}[ht]
\pspicture(0,0)(18.5,20.5)

\rput[tl]{0}(1.,20.){\epsfxsize=17cm
\epsffile{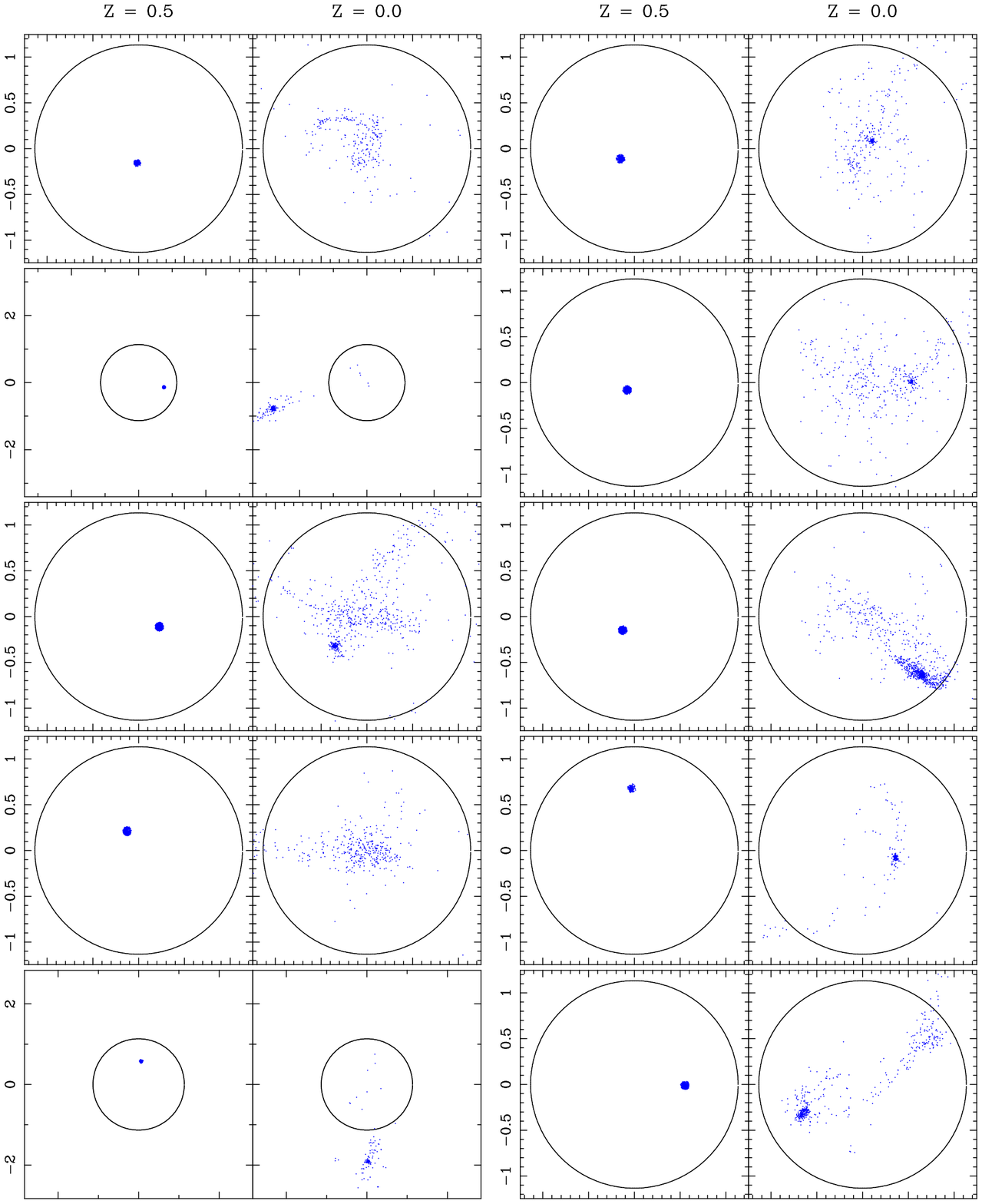}}

\rput[tl]{0}(0.,1.5){
\begin{minipage}{18.5cm}
  \small\parindent=3.5mm {\sc Fig.}~8.--- The same as in Fig.
  {\FigclevolI} but for ten halos randomly selected from the halos
  identified within the virial radius of the cluster at $z=0.5$
  ($\Rvir\approx 1.1\hMpc$ (proper) and $\Mvir\approx 5.8\times
  10^{14}\hMsun$). The first and third columns from the left show the
  positions of the particles bound to the halos at $z=0.5$, while the
  second and the fourth panels show the positions of the same particles
  at $z=0$. Note that despite substantial mass losses, eight out of ten
  of these halos can be identified at $z=0$ as distinct dense clumps of
  particles. The circles in all panels have the same radius equal to
  the virial radius of the cluster at $z=0.5$.
\end{minipage}
}
\endpspicture
\end{figure*}

\begin{figure*}[ht]
\pspicture(0,0)(18.5,19.)

\rput[tl]{0}(1.,19.5){\epsfxsize=17cm
\epsffile{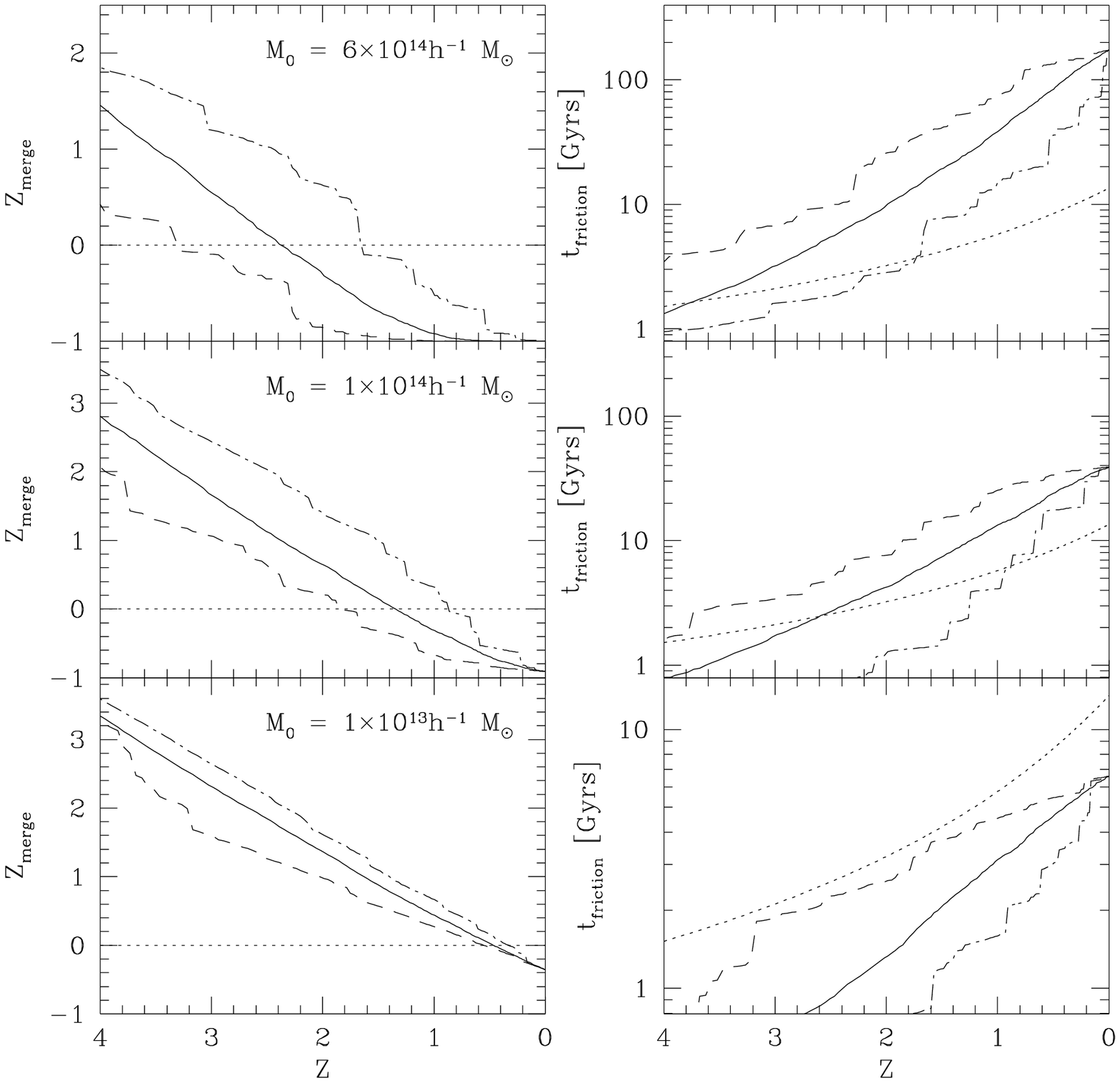}}

\rput[tl]{0}(0.,2.75){
\begin{minipage}{18.5cm}
  \small\parindent=3.5mm {\sc Fig.}~9.--- Estimates of the dynamical
  friction time-scale as a function of time for clusters of different
  final ($z=0$) mass $M_0$.  The right column presents estimates of
  $t_{fric}$ for satellite halos of maximum circular velocity
  $V_{max}=120{\ }{\rm km/s}$. The three curves correspond to the
  average evolution (solid curves), early formation (dashed curves) and
  late formation (dot-dashed curves) computed using Monte-Carlo
  realizations of mass accretion histories (see \S {\Secddnl} for
  details). The dotted curve shows the age of the Universe in the
  {\LCDM} model studied here as a function of redshift.  The left
  column shows the corresponding evolution of ``merging redshift''
  defined as a redshift corresponding to $t(z)+t_{fric}(z)$, where
  $t(z)$ is age of the Universe at redshift $z$ and $t_{fric}$ is the
  corresponding dynamical friction time-scales (both quantities are
  shown in the right panel). The curve marking has the same meaning as
  in the right panel.  Negative values $z_{merger}<0$ mean that the
  dynamical friction time is larger than the time span between redshift
  $z$ and the present.
\end{minipage}
}
\endpspicture
\end{figure*}

The right column of panels in Figure {\Figzmerge} represents the
dynamical friction time in a different, easier to interpret, way. It
shows the ``merging redshift,'' defined as the redshift corresponding to the
time $t+t_{fric}$, where $t$ is the current epoch (redshift $z$). In
the sense of the dynamical friction time, $z_{merge}$ can be
interpreted as a redshift at which most of the halos present in cluster
at redshift $z$, will merge into a central object. This interpretation
assumes that the  mass of the cluster would not change, which is not correct.
The actual value should therefore be considered as an upper limit on
the actual $z_{merge}$.  Negative values $z_{merger}<0$ mean that the
dynamical friction time is larger than the time span between redshift
$z$ and the present.

Figure {\Figzmerge} shows that regardless of the final cluster mass,
dynamical friction is efficient at $z\gtrsim 3$. For small final mass
clusters, dynamical friction is efficient throughout the entire cluster
evolution, while for medium- and high-mass clusters, dynamical
friction effectively switches off as the cluster exceeds a certain mass
threshold. Cluster CL1 in Figure {\Figbpro}, evolves very close to the
average mass evolution track of the $6\times 10^{14}\hMsun$ final mass
clusters. From the estimate of $z_{merge}$ we can therefore expect that
all of its satellite halos will merge into the central object by $z\sim
0.5-1$. Note also that for all cluster masses, only halos accreted
before $z=0.5$ can be substantially influenced by dynamical
friction. However, even at $z<0.5$, dynamical friction may remain
efficient in less massive clusters and groups, which, if accreted by a
cluster, will be depleted of halos and will tend to lower the bias
value further in this cluster.  Both of these results are in
qualitative agreement with simulation (see Figs.
{\FigclevolI},{\FigclevolII}). 

\section{Discussion}

Results and arguments presented in the previous section suggest that we
can identify major processes that drive the evolution of the halo bias.  In
the linear ($\delta_m\lesssim 1$; the overdensities quoted here and
below are for a density field smoothed with the top-hat filter of radius
$R_{\rm TH}=5\hMpc$) and mildly non-linear regimes ($\delta_m\lesssim
3$) the analytical model of bias developed by Mo \& White (1996; see \S
{\Secbanalytic}) is in good agreement with the results of our
simulation, {\em if\/} the formation epoch of halos $z_f$ is distinguished
from the epoch of observation $z$ (or, in other words, if halos are
assumed to retain their identity during a finite interval of time after
their formation).  This model reproduces the non-linearity and
time evolution of the bias observed in the simulation well (see Fig.
{\Figb4plin}). The evolution of the bias in linear and quasi-linear
regimes is, therefore, driven primarily by the halo collapse and
merging rates specific for a given cosmological model.

Interplay between halo formation and the merging rate in high-density
regions (affecting $n(M,z,z_f\vert R_0,\delta_0)$ in eq.[\Eqdh]) and
halo formation and the merging rate in the field (affecting $n(M,z,z_f)$)
leads to the decrease of the bias with time {\em for halos in a given
  mass range}.  This is simply because halos of a given mass collapse
earlier in high-density regions than they do in the field; the
high-redshift objects therefore represent rare events in the density
field and are initially strongly clustered due to modulation by
large-scale modes. The number of halos in high-density regions is then
decreased due to merging, while number density of halos in the field
may still increase (or level off depending on mass range considered) at
lower redshifts. The ratio is thus a decreasing function of time.  In
the regions of negative overdensity ($\delta_m\lesssim -0.5$), the
evolution is reverse: formation of halos of a given mass in these
regions is delayed and at early epochs their number density is below
the average (see Fig. {\Figb4plin}). The halos in underdense regions
are thus anti-biased at early epochs. Figure {\Figb4plin} shows that
bias at $\delta_m\lesssim -0.5$ increases during evolution as the
collapse threshold is reaching lower and lower overdensities and more
and more halos are being formed in these underdense regions (Fig.
{\Figbnonlin} shows that $b\approx 1$ for $-1\lesssim \delta_m\lesssim
2$ at $z=0$). The prediction of the analytical model at these overdensities
matches our numerical results nicely.

In high-density regions ($\delta_m\gtrsim 3$), the evolution and
amplitude of bias appears to be significantly affected by dynamical
friction\footnote{It is clear that dynamical friction {\em is}
  important for halo evolution; indeed, binary halo mergers are also
  due to dynamical friction. However, here we discuss dynamical
  friction that operates on satellites orbiting inside a more massive
  system (a group or a cluster), which leads to the decay of their orbits,
  and, ultimately, to a merger with the central cluster object.}.
Indeed, the process results in halo mergers with the central cluster
halo and reduces thereby the ratio of halo to matter overdensities. The
efficiency of dynamical friction depends sensisitively on the mass of
satellite halos, properties of halo orbits, and mass of the cluster.
The latter evolves rapidly with time and switches dynamical
friction off at some epoch ($z\sim 0.5-1.0$ for the {\LCDM} model
studied here). The mass accretion history is a stochastic process and
some scatter in the mass evolution is expected for clusters that have
the same mass at $z=0$ (Lacey \& Cole 1993): some clusters accrete most
of the mass early on, while others accrete most of their mass at lower
redshifts.  Therefore, for a given observation epoch $z$, different
clusters may be affected by the dynamical friction to a different
degree. For example, a cluster which have accreted most of its mass
(and halos) just prior to $z$, will be less  affected than a cluster of
the same mass that accreted most of its halos earlier because 
dynamical friction had more time to operate in this cluster. The
situation is even more complicated, because clusters, even those in
which dynamical friction becomes inefficient due to mass increase, may
accrete smaller clusters and groups which, in turn, have different
evolution histories and therefore different ratios of halo to matter
overdensities.

Dynamical friction is not the only process that affects halo
counts in clusters and groups. Tidal fields of clusters strip the outer
parts of satellite halos which may result in substantial mass loss
for a medium- and high-mass clusters (up to a factor of $5-20$ depending on
paramaters of the halo orbit and the period of time the halo spends in
cluster; see, e.g., KGKK). The maximum circular velocity, $V_{max}$, of
the halo changes only mildly, even in the case of severe mass loss
(which, as a reminder, is the reason we use it for the halo selection).
Nevertheless, for some halos $V_{max}$ may decrease below the selection
threshold, in which case these halos will ``drop out'' of the catalogs.
This process likely contributes to the mild evolution of bias seen in
Fig. {\Figbpro} at $z\leq 0.5$. The efficiency of tidal stripping is
lower for lower-mass systems; therefore, we expect it to be important
only in relatively massive ($M_{vir}\gtrsim 10^{14}\hMsun$) clusters.
While this effect may seem to depend on the particular halo
selection procedure used in our analysis, similar effects may arise for
other selection procedures\footnote{Note that we are bound to use some
  selection procedure because in most cases we are interested in studying
  the clustering of a particular (selected) class of objects: halos of a
  given type, galaxies of a given luminosity or color, etc.}. It is clear,
for example, that this effect would be even more severe had we chosen to
select halos using their {\em bound} mass. Observationally, 
galaxy catalogs are usually selected using a fixed luminosity limit
in a given wavelength band. If the luminosity of galaxies in this band
evolves differently in clusters than in the field (which is strongly
suggested by a variety of observations), the uniform selection criterion
is bound to select somewhat different galaxy populations in high- and
low-density regions.

Different evolution histories of different systems may result in
different locally evaluated bias. If, for example, matter and halo
density fields are smoothed at some sufficiently large (larger than
typical cluster size) scale, regions of a similar matter
overdensity may correspond to different halo overdensities, because the
latter depends on the evolution history of systems encompassed within
the smoothing scale.  We can expect, therefore, that bias evaluated at
a finite smoothing scale will exhibit some scatter in different regions
of space. Note that this scatter arises not from the ``stochasiticity''
of the halo formation, but from the fact that the same matter
overdensity may correspond to regions of very different evolution
histories and, thus, of different halo content.  On the other hand, the
differences in the evolution histories result from the modulation by
large-scale modes in the density distribution. The bias is thus also
modulated by the large-scale modes and is therefore {\em non-local}.
Note that this ``stochasticity'' should decrease, as one smoothes
density fields on progressively larger scales because effects of the
modulation by large-scale modes are smaller on larger scales. In
general, if the relation between halo and matter density is non-linear, the
bias estimated from a density field smoothed at any particular scale
will be non-linear and will exhibit some scatter (Dekel \& Lahav 1998).
In our simulation, differences in the bias between different regions of
the computational volume seen in Fig.  {\Figbnonlin} are caused by
different numbers and formation histories of clusters and groups in
these regions. The regions that exhibit weaker bias (i.e., stronger
anti-bias) are the regions that contain clusters with earlier formation
epochs. The Coma-size cluster that already had mass of $\approx 1.3\times
10^{13}\hMsun$ at $z=3$ exhibits the highest matter overdensity
and the strongest anti-bias.

Clusters and groups of galaxies contribute a significant fraction of the
galaxy clustering signal at small, $r\lesssim 1-2\hMpc$, scales and
significantly affect clustering amplitude at larger scales. The
anti-bias arising due to dynamical processes in groups and clusters can
therefore explain the anti-bias seen in comparisons of halo and matter
two-point correlation functions and power spectra. Our results then
imply that understanding of the evolution of small-scale galaxy
clustering requires a deeper understanding of the evolution of galaxies in
groups and clusters. 

The major caveat in interpretation of these dissipationless results is
a correspondence between dark matter halos and observable galaxies. We
note, however, that the definition of dark matter halo in this study is
significantly different from a conventional definition, which
eliminates many problems of halo-galaxy mapping related to the
overmerging problem (see KGKK and Col\'{\i}n et al. 1998 for
discussion). In the framework of the hierarchical structure formation
modelled here, it seems likely that in every sufficiently massive
($M\gtrsim 10^{11}\hMsun$) {\em gravitationally bound} halo, at some
epoch baryons will cool, form stars, and produce an object ressembling
a galaxy (e.g., White \& Rees 1978; Kauffman, Nusser \& Steinmetz 1997;
Yepes et al. 1997). This is indeed a cornerstone assumption of the
semi-analytical models of galaxy formation (e.g., Somerville \& Primack
1998).  Therefore, even though we cannot predict detailed properties of
galaxies hosted by dark matter halos (which would require inclusion of
a substantial amount of additional physics in the simulations), it is
likely that the overall halo distribution should be representative of
the expected distribution of the overall galaxy population.  Excellent
agreement between clustering of galaxy-size halos in our simulation and
observed galaxy clustering, demonstrated in comparisons of the
two-point correlation functions (Col\'{\i}n et al. 1998) and power
spectra (see Fig.  {\Figpsz0}), is an indirect but strong indication in
favor of this point.

Anti-bias, similar in amplitude and scale-dependence to that
observed in our simulation, has been found in other recent numerical
studies that employ either non-adiabatic hydrodynamics (Jenkins et al.
1998b) or semi-analytic recipes (Kauffman et al. 1998ab; Diaferio et
al.  1998; Benson et al. 1998) to model galaxy formation and evolution
of the two-point correlation function. These studies are very different in
their modelling technqiues and the agreement indicates that anti-bias
is real and is not a numerical artefact of a particular simulation.
Anti-bias was traditionally an unfavored proposition because it is
easier to envision a biased rather than anti-biased galaxy formation.
However, as we have argued above, the anti-bias may arise naturally
during the dynamical evolution of the halo population, even though halo
formation is biased.  Indeed, all of the numerical studies mentioned
above are similar to the present study in explicit modelling of the
orbital evolution of halos in groups and clusters. Diaferio et al.
(1998) present bias profiles $b(r)$ of the clusters in their simulation
that are in good qualitative agreement with the profiles shown in Fig.
{\Figbpro}, indicating that similar dynamical processes are probably 
causing the anti-bias in their study. 

We can therefore expect that simulations affected by the overmerging
problem and in which a recipe for splitting or weighting the overmerged
halos is used may overestimate the clustering amplitude at small-scales and
not show (or show a weaker) anti-bias. The amount of anti-bias should
also depend on the box size because small-size ($\lesssim 30-50\hMpc$)
boxes are unlikely to contain clusters in the high-mass tail of the
mass function. Such a trend is indeed observed; the comparison of bias in
the {\LCDM} simulations of $30\hMpc$ and $60\hMpc$ boxes presented in
KGKK and Col\'{\i}n et al.  (1998) shows that anti-bias is stronger in
the $60\hMpc$ simulation.  Finally and most importantly, the
sensitivity of the small-scale bias amplitude to the abundance and
evolution histories of clusters indicates that we can expect some
differences between cosmological models. The models that form clusters
at systematically later epochs and/or in lesser abundance, should
exhibit a weaker anti-bias (or even absence of anti-bias), than models
that form clusters earlier and in larger numbers. Thus, for example, a
$\tau$CDM model appears to result in a higher value of $z=0$ bias than
the $\Lambda$CDM model (Kauffman et al. 1998ab, Col\'{\i}n et al. 1998;
Diaferio et al. 1998). A more systematic large-box study is needed,
however, to test and quantify such dependence on cosmology.

We would like to note that the anti-bias (in the amount predicted by
the numerical studies) is actually needed to reconcile the otherwise
very successful {\LCDM} model with observed galaxy clustering at $z=0$
(e.g., Klypin, Primack \& Holtzman 1996; Cole et al. 1997; Jenkins et
al. 1998a). Note that this requirement applies to the overall galaxy
population represented in large galaxy surveys such as the CfA and APM;
sub-populations of galaxies may exhibit significantly different
clustering properties, as is indicated by the differences between
clustering of spiral and elliptical galaxies (e.g. Guzzo et al. 1997),
infrared- and optically-selected galaxies (e.g. Hoyle et al. 1998),
etc. In the simulation presented here, certain sub-samples of halos are
clustered more strongly than the overall population. For example, as
shown in Gottl\"ober, Klypin \& Kravtsov (1998), the population of
halos that loose mass at $z<1$ (the halos that are accreted and being
tidally stripped by clusters) are actually biased at $z=0$ with respect
to the matter distribution, as opposed to the strongly anti-biased
distribution of the entire halo population.  The simulation used in our
analysis was also used by Kolatt et al. (1998), who showed that
interpretation of the high clustering amplitude of the high-redshift
galaxies is not unique. The amplitude can be reproduced equally well
when these galaxies are assumed to be located in high-mass halos, or
are assumed to have smaller masses but undergo a starburst due to a
collision/merger with another galaxy. Recent studies by Blanton et al.
(1998), Cen \& Ostriker (1998), Kauffmann et al. (1998ab), and Diaferio
et al. (1998) demonstrate the existence of age, luminosity, color
segregation of clustering in their models. Some differences in
clustering properties of certain sub-samples from the overall
population can therefore be expected.

Another observational requirement for the {\LCDM} model is a
scale-dependent bias (e.g., Jenkins et al. 1998a): the shape of the observed
galaxy correlation function is rather different from the shape of the 
non-linear matter correlation function; this scale-dependency is also
reproduced nicely in the simulations. Recent studies of galaxy power
spectrum by Gazta\~naga \& Baugh (1998) and Hoyle et al. (1998) stress
that the anti-bias is required at rather low wavenumbers.  However, as
we have noted above (see \S 1) and have illustrated in Fig.
{\Figpsz0}, the amount of small-scale anti-bias observed in the
galaxy-correlation function (Col\'{\i}n et al. 1998) is sufficient to
produce the required anti-bias in the power spectrum at low wavenumbers.
Indeed, the power spectrum of the halo distribution in our simulation
matches nicely the power spectrum of the APM galaxies at all probed
wavenumbers. The matter power spectrum, on the other hand, is different
from the galaxy power spectrum at a very high ($>5-10\sigma$)
significance level.  The existence of non-linear and scale-dependent
bias of the galaxy distribution may affect other analyses that 
either assume there is no bias or that the bias is linear. These include
estimates of the matter density $\Omega$ from peculiar velocities and
redshift distorsions (Dekel \& Lahav 1998; Pen 1998) and from the
observed mass to light ratios in galaxy groups and clusters (Diaferio
et al. 1998), estimates of the baryon density in the Universe (e.g.,
Goldberg \& Strauss 1998; Meiksin, White \& Peacock 1998), estimates of
the cosmological parameters based on the joint analysis of galaxy
redshift surveys, cosmic microwave background anisotropies, and
high-redshift supernovae (e.g, Eisenstein, Hu \& Tegmark \& 1998), etc.
In this respect, the lesson of the present analysis is that any use of
the galaxy density field as a cosmological probe requires very careful
testing and evaluation.

\section{Conclusions}

We presented results from a study of the origin and evolution of bias of
the dark matter halo distribution in a large, high-resolution simulation 
using a low-matter density, flat, CDM model with cosmological constant. The
following conclusions can be drawn from the results presented in this
paper.

1. The evolution of the power spectrum of the halo distribution is
significantly slower than the evolution of the matter power spectrum at all
(both linear and non-linear) scales. The halo and matter power spectra 
also have significantly different shapes. The differences in shape and
rate of evolution imply time- and scale-dependent bias of the halo
distribution which is in qualitative agreement with the results of
the correlation function analyses. We stress, however that the scale
dependence of the bias determined from the power spectrum $b_P(k)\equiv
\sqrt{P_h(k)/P_m(k)}$ is different from the scale dependence of
$b_{\xi}(r)\equiv \sqrt{\xi_h(r)/\xi_m(r)}$, because the two statistics
are related via the Fourier transform (see \S 1). Put simply, $b_P(k)$
cannot be obtained from $b_{\xi}(r)$ by a naive substitution of
variable $r\propto 1/k$. At $z=0$ the halo power spectrum in our
simulation matches the observed power spectrum of the APM
galaxies well.

2. Despite the differences in shape, the power spectra of both matter
and halos exhibit a distinct ``inflection point'' at approximately the
same wavenumber, corresponding to the scale of non-linearity (i.e., the
scale at which the power spectra begin to deviate significantly from
the linear power spectrum). The inflection scale is $\approx
0.15-0.2\ihMpc$ and coincides with the inflection observed in the power
spectrum of APM galaxies (Gazta\~naga \& Baugh 1998); therefore, we
interpret the inflection in the APM power spectrum as the present-day
scale of non-linearity $k_{NL}$. In the simulation analyzed here,
$k_{NL}$ evolves with time from $\sim 1\ihMpc$ at $z=5$ to $\approx
0.15-0.2\ihMpc$ at $z=0$. We should note that the distinct inflection
point can be seen only in the {\em real-space} power spectrum; the
non-linear amplitude of the redshift-space power spectrum is strongly
suppressed and the inflection in the power spectrum of both matter and
halos is smoothed out (see Figs. 3 \& 4 in Gottl\"ober et al. 1998).

3. The analytic fitting formula of Peacock \& Dodds (1996), with only 
minor tuning, provides an excellent match to both the shape and
evolution rate of the matter power spectrum in our simulation. The latter 
probes deep into the non-linear regime, down to wavenumbers of 
$\sim 200\ihMpc$ (at $z=0$); we find that clustering of dark matter 
in the highly non-linear regime in the simulation is approximately 
stationary (stable clustering).  

4. In addition to $b_P$, we examine the evolution of bias defined using
smoothed halo and matter overdensities ($\delta_h$ and $\delta_m$) as
$b_{\delta}\equiv \delta_h/\delta_m$. In agreement with results from
the correlation function and power spectrum analyses, we find that
$b_{\delta}$ is non-linear (i.e., depends on $\delta_m$) and
time-dependent. If we modify the model and assume that halos can retain
their identity for a finite period of time after their formation and
distinguish between formation and observation epochs $z_f$ and $z$, the
analytic model of Mo \& White (1996) can describe both the non-linearity
and evolution of $b_{\delta}$ at linear and quasi-linear overdensities
($\delta_m\lesssim 5-7$, here and below the overdensities are smoothed
with the top-hat smoothing radius of $5\hMpc$). The original model
($z=z_f$) significantly underestimates the bias of galaxy-size halos at
$\delta_m\gtrsim 1$.

5. We argue that at non-linear overdensities the bias evolution is
significantly affected by dynamical friction and tidal stripping of
halos in groups and clusters. Both processes tend to reduce the number
density of cluster halos of a given mass range, thereby reducing the
bias and resulting in anti-bias at $z\lesssim 0.5$ at $\delta_m\gtrsim
5$ in the {\LCDM} model studied here. The effect of dynamical friction
depends sensitively on the cluster formation history, which introduces
a certain degree of scatter into the bias $b_{\delta}$.

In summary, the evolution of the bias of galaxy-size halos observed in the
simulation in linear and quasi-linear regimes can probably be fully
described using the extended Press-Schechter theory. In other words,
the evolution of bias in this regime results from an interplay between
halo formation and merging rates in different regions and in the field.
In the non-linear regime, the halo bias evolution appears to be driven
by the dynamical processes inside clusters and groups. Thus, despite
the apparent complexity, we believe that the origin and evolution of
bias can be understood in terms of the processes that drive the
formation and evolution of dark matter halos and galaxies that they
host: collapse from the density peaks, merging, tidal stripping and
morphological transformation in the high-density regions. Our results
show that {\em these processes may conspire to produce a halo distribution
quite different from the overall distribution of matter}, yet remarkably
similar to the observed distribution of galaxies. This result implies
that detailed modeling of the small-scale galaxy clustering requires
a good understanding of galaxy evolution in clusters. We would like to
emphasize, therefore, the importance of further efforts in modeling
galaxy evolution in both clusters and in the field.

\acknowledgements This work was funded by NSF grant AST-9319970, NASA
grant NAG-5-3842, and NATO grant CRG 972148 to the NMSU.  We would like
to thank Stefan Gottl\"ober for many useful discussions. We are
grateful to Vladimir Avila-Reese and Claudio Firmani for providing us
the Monte-Carlo mass aggregation histories in electronic form, and to
Michael Gross whose routine we have used to calculate the age of the
Universe.  The simulation presented in this paper was done at the
National Center for Supercomputing Applications (NCSA,
Urbana-Champaign, Illinois) and on the Origin2000 computer at the Naval
Research Laboratory.

\end{document}